\documentclass[12pt]{article} 
\usepackage[sectionbib]{natbib}
\usepackage{array,epsfig,fancyhdr,rotating}
\usepackage{hyperref}

\usepackage{todonotes}
\usepackage{amsmath}
\usepackage{amssymb}
\usepackage{amsfonts}
\usepackage{multirow}
\usepackage{amsthm}

\usepackage{mathtools}
\mathtoolsset{showonlyrefs}
\usepackage{booktabs, hhline}

\usepackage{algorithm}
\usepackage[noend]{algpseudocode}

\theoremstyle{definition}

\title{Approximating optimal SMC proposal distributions in individual-based epidemic models}
\author{Lorenzo Rimella, Christopher Jewell and Paul Fearnhead}

\begin{document}
 \maketitle


\begin{abstract}
Many epidemic models are naturally defined as individual-based models: where we track the state of each individual within a susceptible population. Inference for individual-based models is challenging due to the high-dimensional state-space of such models, which increases exponentially with population size. We consider sequential Monte Carlo algorithms for inference for individual-based epidemic models where we make direct observations of the state of a sample of individuals. Standard implementations, such as the bootstrap filter or the auxiliary particle filter are inefficient due to mismatch between the proposal distribution of the state and future observations. We develop new efficient proposal distributions that take account of future observations, leveraging the properties that (i) we can analytically calculate the optimal proposal distribution for a single individual given future observations and the future infection rate of that individual; and (ii) the dynamics of individuals are independent if we condition on their infection rates. Thus we construct estimates of the future infection rate for each individual and then use an independent proposal for the state of each individual given this estimate. Empirical results show order of magnitude improvement in efficiency of the sequential Monte Carlo sampler for both SIS and SEIR models.     
\end{abstract}

\section{Introduction}

 The use of dynamical disease transmission models to inform disease control policy has increased throughout the 21st century for both human and livestock outbreaks, for example, SARS and H1N1 pandemic influenza in humans \cite[]{zhou2004discrete}, avian influenza in poultry \cite[]{van2005quantification}, and foot-and-mouth disease in cloven-hoofed livestock \citep{zhou2004discrete,jewell2009bayesian}.  Most recently, they have been central in informing national-level decisions on social distancing and vaccination strategies for the SARS-CoV-2 pandemic \citep{BrDanEtAl21, FunkEtAl20}.  Besides outbreaks, such models are also useful for studying the dynamics of endemic diseases, with the ability to explain random fluctuations around an otherwise stable case incidence, particularly in highly heterogeneous populations \citep{Britton2010}.
 
 In essence, disease transmission models belong to the class of state transition models, described by a directed (though not necessarily acyclic) graph. For example, the SIS model proposes individuals as existing as ``susceptible'' or ``infected'', and individuals are allowed to transition from either state to the other.   
In a stochastic setting, it is natural to assume that an individual in the population experiences a hazard rate of progressing from some source state to a destination state. 
 This setup has particular relevance for the case when transition hazard rates depend on the individuals' characteristics as well as the characteristics of their relationship with each other. Many applications demand individual-level granularity, particularly when observations are of specific individuals or where disease interventions are targeted to particular individuals \citep{ChapEtAl20, jewell2009interface, Cocker2022}.  
 
 Inference for such models is, however, challenging due to the presence of partial- or total-censoring of transition events, for which the state-space increases exponentially with population size. For example, in an SIS model, we may have noisy observations of which individuals exist in either the S or I states at particular times, but no direct observation of when state transitions occur.

Following \cite{rimella2022}, we consider sequential Monte Carlo (SMC) methods for inference for such models. We show that standard implementations of SMC \cite[]{gordon1993novel,pitt1999filtering} are inefficient for these individual-based epidemic models. In particular, they struggle to propose states for all individuals that will be consistent with future observations. \cite{ju2021sequential} consider how to improve the efficiency of SMC for individual-based epidemic models, but they consider observations of e.g. the number of infected individuals, and their approach does not obviously apply to the observation models we consider.

To improve SMC, we develop a novel proposal distribution that takes account of future observations. The key idea is based on two properties of the dynamics of individual-based epidemic models. First, calculating the conditional distribution of the state of a single individual, given future observations and the future infection rate of the individual is tractable. This can be calculated using standard recursions for finite-state hidden Markov models \cite{rabiner1986introduction} together with the fact that the state-space for a single individual is small (e.g. 2 for an SIS model or 4 for an SEIR model). Second, there is a form of conditional independence across individuals: if we condition on the future infection rates for each individual then the dynamics of the state for individuals are independent of each other. In the models, we consider the infection rate for each individual just depends on the total number of infectious individuals. Thus we can use ideas from \cite{whiteley2021inference} to estimate the future number of infectious individuals. Then conditioning on this estimate, and the corresponding infection rates for each individual, we have a proposal distribution that is independent across individuals, and for each individual is equal to the true conditional distribution of the state given the estimated future infection rates and observations for that individual.

The computational cost of using this proposal is proportional to the number of time-steps at which we have future observations. In practice, we can implement such a proposal distribution just conditioning on future observations over a suitable time window. We show empirically that using this proposal distribution can lead to an order of magnitude improvement in Monte Carlo efficiency, even after accounting for the increased computational cost. 


\section{Preliminaries}

We use bold lowercase letters for vectors, e.g. $\mathbf{a}$, and bold uppercase letters for matrices, e.g. $\mathbf{A}$. We use $\mathbf{A}^{(i,j)}$ for the $(i,j)$-th element of $\mathbf{A}$ and we use $\mathbf{A}^{(i,\bullet)}$ (or $\mathbf{A}^{(\bullet,j)}$) to represent the column vector given by the $(i)$-th column (or the $(j)$-th row) of matrix $\mathbf{A}$. With $\circ$ and $\slash$ we denote the elementwise product and ratio between vectors or matrices. $\mathbf{1}_M$ denotes the $M$-dimensional vector of ones. Given $t,s \in \mathbb{N}$ with $t>s$ we use $[s:t]$ for the set $\{s, \dots, t\}$, which also applies on indexing as a shorthand, e.g. for $t \in \mathbb{N}$ we use $y_{[1:t]}$ for $\{y_1,\dots, y_t\}$. The notation for the main probability distributions is reported in Table \ref{tab:notation}.

\begin{table}[httb!]
    \centering
    \resizebox{14cm}{!}{
    \begin{tabular}{l|llllll}
    \hline
    Distribution & Categorical          & Bernoulli           & Binomial & Gaussian & Uniform & Multinomial \\
    \hline
    Notation & $\mathcal{C}at_M(i|\mathbf{p})$ & $\mathcal{B}e(i|q)$ & $\mathcal{B}in(i|N,q)$ & $\mathcal{N}(a|\mu, \sigma^2)$ & $\mathcal{U}nif(q|a,b)$ & $\mathcal{M}ult(\mathbf{c}|N, \mathbf{p})$\\
    \hline
    \end{tabular}
    }
    \caption{Notation table for probability mass and density functions.}
    \label{tab:notation}
\end{table}


\section{Model}

\subsection{Individual-based epidemic models} \label{subsec:individual_based_models}

In this article we consider individual-based models defined by: the number of compartments $M$, the population size $N$, the initial probability of an individual being assigned to a compartment $(\mathbf{p}_{n,0})_{n \in [1:N]}$ and the probability of an individual to transition from one compartment to the other $(\mathbf{K}_{n,\bullet})_{n \in [1:N]}$, where the stochastic transition matrix $\mathbf{K}_{n,\bullet}$ is defined as a function of an $M$-dimensional vector $\mathbf{c}$, i.e. $\mathbf{c} \to \mathbf{K}_{n,\mathbf{c}}$. In practice, $\mathbf{c}^{(i)}$ is the number of individuals in compartment $i$, and so the transition matrix $\mathbf{K}_{n,\mathbf{c}}$ depends on the compartments' state only, however, more general versions are possible and briefly discussed in Section \ref{sec:discussion} (e.g. spatial models). We use $(\mathbf{x}_t)_{t \geq 0}$ for the population state and $(\mathbf{c}_t)_{t \geq 0}$ for the compartments' state, following:
\begin{itemize}
	\item[Time $0$:] $\mathbf{x}^{(n)}_0 \sim \mathcal{C}at_M(\bullet | \mathbf{p}_{n,0})$ for $n \in [1:N]$ and\\ $\mathbf{c}_0^{(i)} = \sum_{n=1}^N \mathbb{I}_{\mathbf{x}_0^{(n)}}(i)$ for $i \in [1:M]$;
	\item[Time $t$:] $\mathbf{x}^{(n)}_t|\mathbf{x}_{t-1} \sim \mathcal{C}at_M \left ( \bullet \Big{|} \mathbf{K}_{n, \mathbf{c}_{t-1}}^{(\mathbf{x}_{t-1}^{(n)},\bullet)} \right )$ for $n \in [1:N]$ and\\ $\mathbf{c}_t^{(i)} = \sum_{n=1}^N \mathbb{I}_{\mathbf{x}_t^{(n)}}(i)$ for $i \in [1:M]$.
\end{itemize}        

\paragraph{SIS example}
We can make the SIS model heterogeneous by following the construction in \cite{ju2021sequential}. Suppose that we have $d \in \mathbb{N}$ covariates for each individual, we can then define $(\mathbf{w}_n)_{n \in [1:N]}$ as the collection of $d$-dimensional vectors gathering the individual-specific covariates, from which we can compute for $n \in [1:N]$:
\begin{equation*}
	\mathbf{p}_{n,0} = 
	\begin{bmatrix}
	1-\frac{1}{1+\exp{(-\beta_0^{\mathrm{T}} w_n)}} \\ \frac{1}{1+\exp{(-\beta_0^{\mathrm{T}} w_n)}}	
	\end{bmatrix}, \quad \mathbf{K}_{n,c} = 
\begin{bmatrix}
1-\frac{1}{1+\exp{(-\beta_{\lambda}^{\mathrm{T}} w_n)}}\frac{\mathbf{c}^{(2)}}{N} &   \frac{1}{1+\exp{(-\beta_{\lambda}^{\mathrm{T}} w_n)}}\frac{\mathbf{c}^{(2)}}{N}\\
  \frac{1}{1+\exp{(-\beta_{\gamma }^{\mathrm{T}} w_n)}} & 1-\frac{1}{1+\exp{(-\beta_{\gamma }^{\mathrm{T}} w_n)}}
\end{bmatrix}
\end{equation*}
 with $\beta_0 \in \mathbb{R}^{d}$ and $\beta_\lambda,\beta_\gamma \in \mathbb{R}^{d}$. In this model we have individual-specific probabilities of infection and recovery.

\subsection{Observation model}

The observation process is denoted by $(\mathbf{y}_t)_{t \geq 1}$ and given $(\mathbf{q}_{n,t})_{n \in [1:N], t \geq 1}$ with $\mathbf{q}_{n,t} \in [0,1]^M$ we generate observations per each time step $t$ as follows:
\begin{equation} \label{eq:granular_observation}
	\mathbf{y}^{(n)}_t = \mathbf{x}_t^{(n)} \mathbf{r}_t^{(n)} \text{ with } \mathbf{r}_t^{(n)}\sim \mathcal{B}e \left (\bullet \Big{|} \mathbf{q}_{n,t}^{(\mathbf{x}_t^{(n)})} \right ) \text{ for } n \in [1:N],
\end{equation}
which we refer to as the ``granular observations model''. Note that $\mathbf{y}^{(n)}_t \in [0:M]$ meaning that we either report the state of individual $n$ as it is ($\mathbf{y}^{(n)}_t = \mathbf{x}^{(n)}_t$) or we do not report it at all ($\mathbf{y}^{(n)}_t=0$). This model includes observations from random samples of the population, where each component of $\mathbf{q}_{n,t}$ is the same and equal to the probability that individual $n$ is included in the sample at time $t$, as well as situations where observations are preferentially made for certain states (such as observations being of infected farms for foot-and-mouth disease).
To simplify the notation and derivations, in this paper we focus on individual homogeneous reporting rates, i.e. $\mathbf{q}_{n,t} = \mathbf{q}_t$, and under this assumption $\sum_{n \in [1:N]} \mathbb{I}_{\mathbf{y}^{(n)}_t}(i) \sim \mathcal{B}in(\bullet|\mathbf{c}_t^{(i)}, \mathbf{q}_t^{(i)})$ for any $i \in [1:M]$, which recover the binomial observation model \citep{whiteley2021inference,ju2021sequential}. 

\paragraph{SIS example}
Given $\mathbf{q}_t \in [0,1]^2$, $\mathbf{q}_t^{(1)}$ is the probability of reporting a susceptible, while $\mathbf{q}_t^{(2)}$ is the probability of reporting an infected. 

\subsection{Inference in individual-based models with granular observations} \label{subsec:inference}
In epidemiology, we are interested in inferring both the unknown state of the population $\mathbf{x}_t$ and the parameters of the epidemic $\theta$. Given the time horizon $t$, the individual-based model with granular observation $(\mathbf{x}_s, \mathbf{y}_s)_{s \in [1:t]}$ is by construction a hidden Markov model (HMM). We can hence compute filtering distribution $p(\mathbf{x}_s|\mathbf{y}_{[1:s]},\theta)$ and marginal likelihood $p(\mathbf{y}_{[1:s]}|\theta)$ with the forward algorithm \citep{rabiner1986introduction}. The parameters can be then inferred through, for example, the EM algorithm \citep{yang2017statistical}. 

The forward algorithm requires marginalizing over the whole state-space, making it unfeasible for our individual-based model, where marginalizations are $\mathcal{O}(M^N)$. Alternatively, Sequential Monte Carlo (SMC) algorithms can be employed to obtain particle approximations of $p(\mathbf{x}_s|\mathbf{y}_{[1:s]},\theta)$ and $p(\mathbf{y}_{[1:s]}|\theta)$ \citep{ionides2006inference,kucharski2020early} at a cost that is linear in the number of particles and time horizon. Given a  number of particles $P \in \mathbb{N}$, at each time step $s$ an SMC algorithm proposes instances $(\mathbf{x}_s^p)_{p \in [1:P]}$ of the latent process $(\mathbf{x}_s)_{s \in [1:t]}$ through the proposal distribution $q(\mathbf{x}_s|\mathbf{x}_{s-1}, \mathbf{y}_{[1:t]})$, with $q(\mathbf{x}_0|\mathbf{x}_{-1}, \mathbf{y}_{[1:t]}) \coloneqq q(\mathbf{x}_0|\mathbf{y}_{[1:t]})$ and $q(\mathbf{x}_0|\mathbf{y}_{[1:t]})$ proposal distribution at time $s=0$, and it assigns weights $(w_s^p)_{p \in [1:P]}$ to the particles to produce an importance sample that approximates the filtering distribution. Before moving to the next step, the algorithm uses a resampling scheme $r_s(i)$, a distribution over the particles' indexes $[1:P]$, to discard low-weight particles. At the end of the procedure, particle estimates of the filtering distribution, $p(\mathbf{x}_s|\mathbf{y}_{[1:s]},\theta) \approx (P)^{-1} \sum_{p\in [1:P]} {w}^p_s \delta_{\mathbf{x}_s^p}(\mathbf{x}_s)$, and the marginal likelihood, $p(\mathbf{y}_{[1:t]}|\theta) \approx \prod_{s \in [1:t]}(P)^{-1} \sum_{p\in [1:P]} {w}^p_s$, are generated.

\begin{figure}[httb!]
    \centering
    \includegraphics[width= 0.8\textwidth]{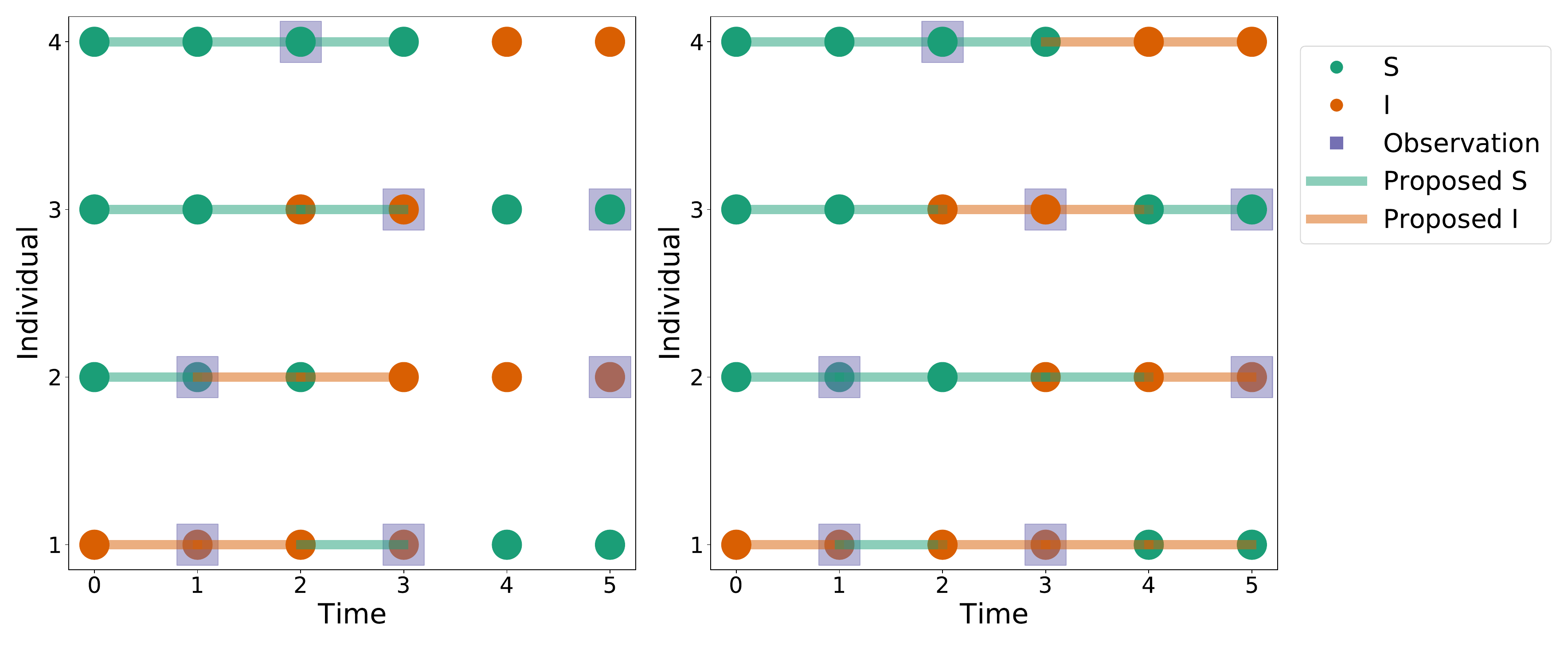}
    \caption{Illustration of BPF (left) and APF (right), in an SIS scenario. Colored dots show the state of each individual, with green for susceptible and red for infected. Dots in grey squares are observations. Horizontal lines from $s-1$ to $s$ are used for the proposed states in $s$.}
    \label{fig:BPF_APF_illustration}
\end{figure}

The performance of SMC algorithms heavily depends on the proposal distributions $(q(\mathbf{x}_s|\mathbf{x}_{s-1}, \mathbf{y}_{[1:t]}))_{s \in [0:t]}$ and the resampling scheme $(r_s(i))_{s \in [0:t]}$, incautious choices of these quantities might lead to high variance of the marginal likelihood estimator, particles\slash weights degeneracy and even observation mismatch, which might cause the failure of the algorithm. The Bootstrap Particle Filter (BPF) \citep{gordon1993novel,candy2007bootstrap} proposes new particles through the transition kernel and it resamples according to the current weights, i.e. $q(\mathbf{x}_s|\mathbf{x}_{s-1}, \mathbf{y}_{[1:t]}) = p(\mathbf{x}_s|\mathbf{x}_{s-1},\theta)$, with $q(\mathbf{x}_0|\mathbf{y}_{[1:t]}) = p(\mathbf{x}_0|\theta)$, and $r_s(i) = \mathcal{C}at_{P}\left (i|\left [{w}_s^1, \dots, {w}_s^P \right ] \right )$. BPF is known to perform poorly in high-dimensional scenarios \citep{bickel2008sharp} and with informative observation, especially when simulated particles have to match certain paths. An easy fix is to include the information from the current observations in the proposal distributions, to avoid mismatch at the current time step when proposing new particles. The resulting algorithm is called the auxiliary particle filter (APF) \citep{pitt1999filtering,carpenter1999improved,johansen2008note} and it arises by picking $q(\mathbf{x}_s|\mathbf{x}_{s-1}, \mathbf{y}_{[1:t]}) = p(\mathbf{x}_s|\mathbf{x}_{s-1}, \mathbf{y}_{s},\theta)$, with $q(\mathbf{x}_0|\mathbf{y}_{[1:t]}) = p(\mathbf{x}_0|\theta)$, and $r_s(i) = \mathcal{C}at_{P}\left (i|\left [{w}_s^1, \dots, {w}_s^P \right ] \right )$. We illustrate graphically in Figure \ref{fig:BPF_APF_illustration} the comparison between BPF and APF in an individual-based model. The BPF fails after three iterations because the proposed particle mismatch the observed state for individuals $1$ and $3$, indeed we observe $\mathbf{y}_3^{(1)}=2$ and $\mathbf{y}_3^{(3)}=2$, but the BPF proposes $(\mathbf{x}^p_3)^{(1)}=1$ and $(\mathbf{x}^p_3)^{(3)}=1$ (green lines). On the contrary, the APF is able to propose particles that are constrained to match the observation, because it includes the current data in the proposal. However, the APF's proposal is still inefficient as it does not take into account future observations. In Figure \ref{fig:BPF_APF_illustration} this is seen by it tending to propose a switch to the infected state immediately before the observation of an infected individual, whereas often an individual becomes infected one or more time-steps earlier. For more complicated models, such as the SEIR model we consider in Section \ref{subsec:SEIR}, the APF can also suffer from mismatch, as the transition to an observed state may not be possible for the current state of a particle.

In the next section, we show how to build for any $s \in [0:t]$ an approximation of $p(\mathbf{x}_s|\mathbf{x}_{s-1}, \mathbf{y}_{[s:t]},\theta)$ for the individual-based model with granular observations. Given that computing $p(\mathbf{x}_s|\mathbf{x}_{s-1}, \mathbf{y}_{[s:t]},\theta)$ requires $(\mathbf{c}_{\tilde{s}})_{\tilde{s} \in [s:t-1]}$, the main idea consists of approximating $(\mathbf{c}_{\tilde{s}})_{\tilde{s} \in [s:t-1]}$ with the expectation of a precomputed multinomial distribution \citep{whiteley2021inference} and to propagate backward the observation $\mathbf{y}_{[s:t]}$ to inform the proposal in $s$.

\section{Optimal proposal distributions for individual-based models}
 
The optimal proposal for an SMC is $p(\mathbf{x}_s|\mathbf{x}_{s-1}, \mathbf{y}_{[s:t]},\theta)$ and it can be compute recursively:
\begin{itemize}
	\item[Time $t$:] $p(\mathbf{y}_t|\mathbf{x}_{t-1},\theta) = \sum_{\mathbf{x}_t \in [1:M]^N} p(\mathbf{y}_t|\mathbf{x}_t,\theta)p(\mathbf{x}_t|\mathbf{x}_{t-1},\theta)$ and \\
	$p(\mathbf{x}_t|\mathbf{x}_{t-1},\mathbf{y}_{t},\theta) = \frac{p(\mathbf{y}_t|\mathbf{x}_t,\theta) p(\mathbf{x}_t|\mathbf{x}_{t-1},\theta)}{p(\mathbf{y}_t|\mathbf{x}_{t-1},\theta)}$;
	\item[Time $s$:] $p(\mathbf{y}_{[s:t]}|\mathbf{x}_{s-1},\theta) = \sum_{\mathbf{x}_s \in [1:M]^N} p(\mathbf{y}_{[s+1:t]}|\mathbf{x}_s,\theta)p(\mathbf{y}_s|\mathbf{x}_s,\theta) p(\mathbf{x}_s|\mathbf{x}_{s-1},\theta)$ and\\
	$p(\mathbf{x}_s|\mathbf{x}_{s-1},\mathbf{y}_{[s:t]},\theta) =  \frac{p(\mathbf{y}_{[s+1:t]}|\mathbf{x}_s,\theta)p(\mathbf{y}_s|\mathbf{x}_s,\theta) p(\mathbf{x}_s|\mathbf{x}_{s-1},\theta)}{p(\mathbf{y}_{[s:t]}|\mathbf{x}_{s-1},\theta)}$;
	\item[Time $0$:] $p(\mathbf{y}_{[1:t]},\theta) = \sum_{\mathbf{x}_0 \in [1:M]^N} p(\mathbf{y}_{[1:t]}|\mathbf{x}_0,\theta) p(\mathbf{x}_0,\theta)$ and\\
	$p(\mathbf{x}_0|\mathbf{y}_{[1:t]},\theta) =  \frac{p(\mathbf{y}_{[1:t]}|\mathbf{x}_0,\theta)p(\mathbf{x}_0,\theta)}{p(\mathbf{y}_{[1:t]}|\theta)}$.
\end{itemize}

See \cite{fearnhead2008computational} for a review of the optimal proposal for importance sampling, \cite{chopin2020introduction} for a discussion on optimal proposal for Particle Filters and \cite{whiteley2014twisted} for a more technical discussion.

A marginalization over the whole state-space is required resulting in a computational cost of $\mathcal{O}(M^N)$ per each step. Observe that at the beginning of the recursion we can exploit the factorization over the individuals at time $t$, components of $\mathbf{x}_t$, of the transition kernel and emission distribution to reduce the computational cost of the marginalization to $\mathcal{O}(N M)$:
\begin{equation}\label{eq:factorization_initial_step}
\begin{split}
	p(\mathbf{y}_t|\mathbf{x}_{t-1},\theta) 
	&= 
	\prod_{n \in [1:N]} \sum_{\mathbf{x}^{(n)}_t \in [1:M]} \mathbf{K}_{n, \mathbf{c}_{t-1}}^{(\mathbf{x}_{t-1}^{(n)},\mathbf{x}_t^{(n)})} \left ( \mathbf{q}_t^{(\mathbf{x}_t^{(n)})} \right )^{\mathbb{I}_{\mathbf{y}_t^{(n)}}(\mathbf{x}_t^{(n)})} \left ( 1- \mathbf{q}_t^{(\mathbf{x}_t^{(n)})} \right )^{\mathbb{I}_{\mathbf{y}_t^{(n)}}(0)}.
\end{split}
\end{equation} 
Notice that $\mathbf{y}_t^{(n)}$ is not conditionally independent given $\mathbf{x}_{t-1}^{(n)}$ because of the dependence of the transition kernel on the compartments' state $\mathbf{c}_{t-1}$. This breaks the computational trick because we cannot express $p(\mathbf{y}_t|\mathbf{x}_{t-1},\theta)$ as a product over the individuals at time $t-1$, and so the cheap marginalization has to be repeated for each state of $\mathbf{x}_{t-1}$, leading to $\mathcal{O}( N M^{N+1})$. However, $\mathbf{y}_t^{(n)}$ is conditionally independent given $\mathbf{x}_{t-1}^{(n)}$ and $\mathbf{c}_{t-1}$, meaning that if an estimate of $\mathbf{c}_{t-1}$ is available a priori the factorization is preserved and the same trick can be iterated in the next time steps.

\subsection{A priori estimates of the compartments' states}\label{sec:whiteley_rime_app}
\cite{whiteley2021inference} proposes an efficient way to approximate the smoothing distribution $p(\mathbf{c}_s|\mathbf{y}_{[1:t]},\theta)$ with a multinomial distribution $\mathcal{M}ult(\mathbf{c}_s|N, \mathbf{m}_{s|t})$ whose parameters are computed recursively with a forward and a backward step through the data at a computational cost $\mathcal{O}(t M^3)$. In the multinomial approximation, there are two key assumptions: the homogeneity of the individuals and a binomial observation model of the form $\mathcal{B}in(\bullet|\mathbf{c}_s^{(i)}, \mathbf{q}_s^{(i)})$. We can recover homogeneity in the individual-based model with granular observation by defining the mean initial distribution $\mathbf{\bar{p}}_{n,0}$ and the mean transition kernel $\mathbf{\bar{K}}_{\mathbf{c}_s}$:  
\begin{equation}
\begin{split}
	&\mathbf{\bar{p}}_{n,0}^{(i)} \coloneqq \frac{1}{N}\sum_{n \in [1:N]} \mathbf{p}_{n,0}^{(i)}, \text{ for } i \in [1:M],\\
	&\mathbf{\bar{K}}_{\mathbf{c}_s}^{(i,j)} \coloneqq \frac{1}{N}\sum_{n \in [1:N]} \mathbf{K}_{n,\mathbf{c}_s}^{(i,j)}, \text{ for } i,j \in [1:M].\\
\end{split}
\end{equation}
We remark that recovering homogeneity by approximating the individuals' transition kernel with an average is also a key step in \cite{ju2021sequential}, where the transition probabilities are approximated by averaging over the individuals to avoid an exponential computational cost in the population size.

We already have $\sum_{n \in [1:N]} \mathbb{I}_{\mathbf{y}^{(n)}_s}(i) \sim \mathcal{B}in(\bullet|\mathbf{c}_s^{(i)}, \mathbf{q}_s^{(i)})$ for $i \in [1:M]$, from which we can define the cumulative observations per each compartment as the vector $\mathbf{o}_s$ with components $\mathbf{o}_s^{(i)} \coloneqq \sum_{n \in [1:N]} \mathbb{I}_{\mathbf{y}^{(n)}_s}(i)$. 

Using the aforementioned approximate dynamic and observation model, the multinomial approximation in \cite{whiteley2021inference} scans the data forward and backward and computes multinomial approximations of the filtering and smoothing distribution (the full algorithm is reported in the appendix). The forward pass consists of a prediction step and an update step preserving the multinomial form, precisely, starting from $\mathbf{m}_{0|0} \coloneqq \mathbf{\bar{p}}_{n,0}$, we have:
$$
\mathbf{m}_{s-1|s} \coloneqq 
\left (\mathbf{m}_{s-1|s-1}^{\mathrm{T}} \mathbf{\bar{K}}_{\mathbf{m}_{s-1}} \right )^{\mathrm{T}}, \quad
\mathbf{m}_{s|s} \coloneqq \frac{\mathbf{o}_s}{N} + \left (  1 - \frac{\mathbf{1}_M^{\mathrm{T}}\mathbf{o}_s}{N}\right ) \frac{\mathbf{m}_{s-1|s} \circ (\mathbf{1}_M-\mathbf{q}_s)}{1-\mathbf{m}_{s-1|s}^{\mathrm{T} }\mathbf{q}_s},
$$
which gives an approximation for the filtering distribution $p(\mathbf{c}_s|\mathbf{y}_{[1:s]}) \approx \mathcal{M}ulti(\mathbf{c}_s| N, \mathbf{m}_{s|s})$. The backward pass implements a reverse kernel and applies it backward:
$$
\mathbf{L}_s \coloneqq \left \{ \left [ (\mathbf{m}_{s|t} \mathbf{1}_M^{\mathrm{T}}) \circ \mathbf{\bar{K}}_{\mathbf{m}_s} \right] \slash \left [ \mathbf{1}_M (\mathbf{m}_{s|t}^{\mathrm{T}} \mathbf{\bar{K}}_{\mathbf{m}_{s}}) \right ] \right \}^{\mathrm{T}}, \quad
\mathbf{m}_{s|t} \coloneqq \left ( \mathbf{m}_{s+1|T}^{\mathrm{T}} \mathbf{L}_s \right )^{\mathrm{T}},
$$
outputting the $M$-dimensional probability vector $\mathbf{m}_{s|t}$ and so approximating the smoothing distribution with $p(\mathbf{c}_s|\mathbf{y}_{[1:t]}, \theta) \approx \mathcal{M}ulti(\mathbf{c}_s|N, \mathbf{m}_{s|t})$. Given the multinomial approximations we can approximate the compartments' state with:
\begin{equation} \label{eq:multinomial_approx_compartment}
	\mathbf{c}_s \approx \mathbb{E}_{\mathcal{M}ult(\mathbf{c}_s|N,\mathbf{m}_{s|t})}(\mathbf{c}_s) = N \mathbf{m}_{s|t}.
\end{equation}

We have imposed a restriction on the emission distribution by assuming a uniform reporting probability for all individuals. However, our approach can be extended to accommodate a more general scenario where $\mathbf{q}_{t,n}$ varies with $n$. To do so, we can compute the mean reporting rate $\bar{\mathbf{q}}_{t} \coloneqq (N)^{-1} \sum_{n \in [1:N]} \mathbf{q}_{t,n}$ when running \cite{whiteley2021inference} and then substitute back $\mathbf{q}_{t,n}$ when computing the approximation to the optimal proposal.

\subsection{Approximate optimal proposals for individual-based models}
Conditioning on $\mathbf{c}_{\tilde{s}} = N \mathbf{m}_{\tilde{s}|t}$ for $\tilde{s} \in [s:t]$ makes the individuals evolve independently from each other and so it allows an analytical computation of $p(\mathbf{y}_{[s:t]}|\mathbf{x}_{s-1},\theta)$ at a cost $\mathcal{O}(N M)$. Starting again from \eqref{eq:factorization_initial_step}:
\begin{equation}\label{eq:factorization_initial_step_approx}
\begin{split}
p(\mathbf{y}_t|\mathbf{x}_{t-1},\theta) &\approx
\prod_{n \in [1:N]} \sum_{\mathbf{x}^{(n)}_t \in [1:M]} \mathbf{K}_{n, N \mathbf{m}_{t-1|t}}^{(\mathbf{x}_{t-1}^{(n)},\mathbf{x}_t^{(n)})} \left ( \mathbf{q}_t^{(\mathbf{x}_t^{(n)})} \right )^{\mathbb{I}_{\mathbf{y}_t^{(n)}}(\mathbf{x}_t^{(n)})} \left ( 1- \mathbf{q}_t^{(\mathbf{x}_t^{(n)})} \right )^{\mathbb{I}_{\mathbf{y}_t^{(n)}}(0)}\\
&\eqqcolon \prod_{n \in [1:N]} \boldsymbol{\xi}_{n,t-1}^{(\mathbf{x}_{t-1}^{(n)})},
\end{split}
\end{equation} 
where we define the quantities $\boldsymbol{\xi}_{n,t-1}$ for each individual $n$ as the approximate probability of observing the future observation $\mathbf{y}_t^{(n)}$ given the state at time $t-1$. We can then follow a similar argument and approximate $p(\mathbf{y}_{[s:t]}|\mathbf{x}_{s-1},\theta)$ as follows:
\begin{equation}\label{eq:factorization_general_step_approx}
	\begin{split}
		p(\mathbf{y}_{[s:t]}|\mathbf{x}_{s-1},\theta) &\approx
		\prod_{n \in [1:N]} \sum_{\mathbf{x}^{(n)}_s \in [1:M]} \boldsymbol{\xi}_{n,s}^{(\mathbf{x}_{s}^{(n)})} \mathbf{K}_{n, N \mathbf{m}_{s-1|t}}^{(\mathbf{x}_{s-1}^{(n)},\mathbf{x}_s^{(n)})} \left ( \mathbf{q}_s^{(\mathbf{x}_s^{(n)})} \right )^{\mathbb{I}_{\mathbf{y}_s^{(n)}}(\mathbf{x}_s^{(n)})} \left ( 1- \mathbf{q}_s^{(\mathbf{x}_t^{(n)})} \right )^{\mathbb{I}_{\mathbf{y}_s^{(n)}}(0)}\\
		&\eqqcolon \prod_{n \in [1:N]} \boldsymbol{\xi}_{n,s-1}^{(\mathbf{x}_{s-1}^{(n)})},
	\end{split}
\end{equation} 
where $\boldsymbol{\xi}_{n,s-1}$ is the approximate probability for each individual $n$ of observing the future observation $\mathbf{y}_{[s:t]}$ given the state at time $s-1$. Note that the marginalization is repeated for all the states of $\mathbf{x}_{s-1}^{(n)}$ and not $\mathbf{x}_{s-1}$, which reduces the cost from $\mathcal{O}(N M^{N+1})$ to $\mathcal{O}(N M^2)$. We can now build our proposal distribution for SMC and approximate $p(\mathbf{x}_s|\mathbf{x}_{s-1}, \mathbf{y}_{[s:t]},\theta)$ as:
\begin{equation}\label{eq:approximate_optimal_proposal}
	\begin{split}
		 p(\mathbf{x}_s|\mathbf{x}_{s-1},\mathbf{y}_{[s:t]},\theta) &\approx
		\prod_{n \in [1:N]} 
		\frac{\boldsymbol{\xi}_{n,s}^{(\mathbf{x}_{s}^{(n)})} \mathbf{K}_{n,  \mathbf{c}_{s-1}}^{(\mathbf{x}_{s-1}^{(n)},\mathbf{x}_s^{(n)})} \left ( \mathbf{q}_s^{(\mathbf{x}_s^{(n)})} \right )^{\mathbb{I}_{\mathbf{y}_s^{(n)}}(\mathbf{x}_s^{(n)})} \left ( 1- \mathbf{q}_s^{(\mathbf{x}_s^{(n)})} \right )^{\mathbb{I}_{\mathbf{y}_s^{(n)}}(0)}}{\boldsymbol{\tilde{\xi}}_{n,s}^{(\mathbf{x}_{s-1}^{(n)})}},\\
		 p(\mathbf{x}_0|\mathbf{y}_{[1:t]},\theta) &\approx
		\prod_{n \in [1:N]} 
		\frac{\boldsymbol{\xi}_{n,0}^{(\mathbf{x}_{0}^{(n)})} \mathbf{p}_{n,0}^{(\mathbf{x}_0^{(n)})}}{\tilde{{\xi}}_{n,0}},
	\end{split}	
\end{equation}
for $s \in [1:t]$ and with:
\begin{equation} 
\begin{split}
	&\boldsymbol{\tilde{\xi}}_{n,s}^{(\mathbf{x}_{s-1}^{(n)})} \coloneqq \sum_{\mathbf{x}^{(n)}_s \in [1:M]} \boldsymbol{\xi}_{n,s}^{(\mathbf{x}_{s}^{(n)})} \mathbf{K}_{n,  \mathbf{c}_{s-1}}^{(\mathbf{x}_{s-1}^{(n)},\mathbf{x}_s^{(n)})} \left ( \mathbf{q}_s^{(\mathbf{x}_s^{(n)})} \right )^{\mathbb{I}_{\mathbf{y}_s^{(n)}}(\mathbf{x}_s^{(n)})} \left ( 1- \mathbf{q}_s^{(\mathbf{x}_s^{(n)})} \right )^{\mathbb{I}_{\mathbf{y}_s^{(n)}}(0)},\\
	&\tilde{{\xi}}_{n,0} \coloneqq \sum_{\mathbf{x}^{(n)}_0 \in [1:M]} \boldsymbol{\xi}_{n,0}^{(\mathbf{x}_{0}^{(n)})} \mathbf{p}_{n,0}^{(\mathbf{x}_0^{(n)})},
\end{split}
\end{equation}
for $n \in [1:N]$. 

It is crucial to understand the difference between ${\boldsymbol{\xi}}_{n,s}$ and $\boldsymbol{\tilde{\xi}}_{n,s}$. ${\boldsymbol{\xi}}_{n,s}$ is used to approximate $p(\mathbf{y}_{[s+1:t]}|\mathbf{x}_{s},\theta)$ without knowing $\mathbf{x}_{s}$ and so it is computed by substituting $\mathbf{c}_{s}$ with $N \mathbf{m}_{s|t}$.  $\boldsymbol{\tilde{\xi}}_{n,s}$ is used to approximate $p(\mathbf{y}_{[s+1:t]}|\mathbf{x}_{s},\theta)$ when knowing $\mathbf{x}_{s}$ and so having access to the actual $\mathbf{c}_{s}$. The latter is important because when considering the proposal distribution of an SMC we know the latest particles and we want to propose the next time step given the last. It is worth mentioning the special case $s=0$, here we have no latest particles hence the recursion looks different, in particular, $\tilde{{\xi}}_{n,0}$ is a scalar and it can be used to approximate the marginal likelihood $p(\mathbf{y}_{[1:t]}|\theta)$. Note that the marginal likelihood approximation could be a useful tool, for example, it can be employed in pseudo-likelihood methods \citep{andrieu2009pseudo} or implemented in a delayed acceptance particle MCMC \citep{golightly2015delayed}.

\begin{algorithm}[httb!]
	\caption{Computation of $({\boldsymbol{\xi}}_{n,h,s},\boldsymbol{\tilde{\xi}}_{n,h,s})_{n \in [1:N]}$} \label{alg:alternative_proposal_h}
	\begin{algorithmic}[1]
		\small
		\Require{$(\mathbf{K}_{n,\bullet})_{n \in [1:N]}$, $(\mathbf{q}_{\tilde{s}})_{\tilde{s} \in [s+1:s+h]}$, $(\mathbf{m}_{\tilde{s}|t})_{\tilde{s} \in [s:s+h-1]}$, $\mathbf{y}_{[s+1:s+h]}$, \qquad \text{\textbf{if} $s\neq 0$} \text{ \textbf{add} $\mathbf{y}_s, \mathbf{c}_{s-1}$}}
		\For{$n=1,\dots,N$}
		\State $\boldsymbol{\xi}_{n,h,s+h}\leftarrow \mathbf{1}_M$
		\For{$\tilde{s} = s+h-1,\dots, s$}
		\State $\boldsymbol{\xi}_{n,h,\tilde{s}}^{\mathrm{T}} \leftarrow \mathbf{K}_{n, N \mathbf{m}_{\tilde{s}|t}}^{(\bullet,i)} \mathbf{q}_{\tilde{s}+1}^{(i)}\boldsymbol{\xi}_{n,h,\tilde{s}+1}^{(i)} \mathbb{I}_{\mathbf{y}_{\tilde{s}+1}^{(n)}}(i) + \mathbf{K}_{n, N \mathbf{m}_{\tilde{s}|t}}( \mathbf{1}_M- \mathbf{q}_{\tilde{s}+1} \circ \boldsymbol{\xi}_{n,h,\tilde{s}+1} ) \mathbb{I}_{\mathbf{y}_{\tilde{s}+1}^{(n)}}(0)$
		\EndFor
		\If{$s \neq 0$}
		\State $\boldsymbol{\tilde{\xi}}_{n,h,s}^{\mathrm{T}} \leftarrow \mathbf{K}_{n,  \mathbf{c}_{s-1}}^{(\bullet,i)}\mathbf{q}_s^{(i)}\boldsymbol{\xi}_{n,h,s}^{(i)}\mathbb{I}_{\mathbf{y}_{s}^{(n)}}(i) + \mathbf{K}_{n,  \mathbf{c}_{s-1}}\left ( \mathbf{1}_M- \mathbf{q}_s \circ \boldsymbol{\xi}_{n,h,s} \right )\mathbb{I}_{\mathbf{y}_{s}^{(n)}}(0)$
		\Else
		\State $\tilde{{\xi}}_{n,h,0} \leftarrow \mathbf{p}_{0,n}^{\mathrm{T}}  \boldsymbol{\xi}_{n,h,1}$
		\EndIf
		\EndFor
	\end{algorithmic}
\end{algorithm} 

${\boldsymbol{\xi}}_{n,s}$ and $\boldsymbol{\tilde{\xi}}_{n,s}$ are the only quantities needed to compute our approximate proposal distribution and they can be precomputed before running the SMC at a computational cost $\mathcal{O}(t N M^2)$. However, this also requires a memory cost of $\mathcal{O}(t N M^2)$, because they have to be accessible when running the SMC and it is a considerable issue when $t$ is large. As an alternative, we can compute ${\boldsymbol{\xi}}_{n,s}$ and $\boldsymbol{\tilde{\xi}}_{n,s}$ at each step of the SMC, which requires a computational cost of $\mathcal{O}(P t^2 N M^2)$. A quadratic in $t$ computational cost is still undesirable, hence we can reduce it by using the observations from the closest future instead of the whole sequence. We can indeed focus on approximating $p(\mathbf{x}_s|\mathbf{x}_{s-1},\mathbf{y}_{[s:s+h]},\theta)$ for $h \in \mathbb{N}$ and $h \ll t$. Given that we have presented our approximation for an arbitrary $t$, approximating $p(\mathbf{x}_s|\mathbf{x}_{s-1},\mathbf{y}_{[s:s+h]},\theta)$ is like approximating $p(\mathbf{x}_s|\mathbf{x}_{s-1},\mathbf{y}_{[s:t]},\theta)$ for $t = s+h$, but we make the dependence on $h$ explicit by defining ${\boldsymbol{\xi}}_{n,h,s}$ and $\boldsymbol{\tilde{\xi}}_{n,h,s}$ as the ${\boldsymbol{\xi}}_{n,s}, \boldsymbol{\tilde{\xi}}_{n,s}$ obtained from the algorithm when looking $h$ steps ahead. The whole procedure is summarized in Algorithm \ref{alg:alternative_proposal_h} and it requires a computational cost of $\mathcal{O}(h N M^2)$. Embedding this algorithm in an SMC demands a computational cost of $\mathcal{O}(P t h N M^2)$, which can be controlled by the users depending on the computational resources and application. We can then conclude the section by stating our optimal proposal distribution:
\begin{equation}\label{eq:approximate_optimal_proposal_q}
\begin{split}
q(\mathbf{x}_s|\mathbf{x}_{s-1},\mathbf{y}_{[1:t]},\theta) &=
\prod_{n \in [1:N]} 
\frac{\boldsymbol{\xi}_{n,h,s}^{(\mathbf{x}_{s}^{(n)})} \mathbf{K}_{n,  \mathbf{c}_{s-1}}^{(\mathbf{x}_{s-1}^{(n)},\mathbf{x}_s^{(n)})} \left ( \mathbf{q}_s^{(\mathbf{x}_s^{(n)})} \right )^{\mathbb{I}_{\mathbf{y}_s^{(n)}}(\mathbf{x}_s^{(n)})} \left ( 1- \mathbf{q}_s^{(\mathbf{x}_s^{(n)})} \right )^{\mathbb{I}_{\mathbf{y}_s^{(n)}}(0)}}{\boldsymbol{\tilde{\xi}}_{n,h,s}^{(\mathbf{x}_{s-1}^{(n)})}},\\
q(\mathbf{x}_0|\mathbf{y}_{[1:t]},\theta) &=
\prod_{n \in [1:N]} 
\frac{\boldsymbol{\xi}_{n,h,0}^{(\mathbf{x}_{0}^{(n)})} \mathbf{p}_{n,0}^{(\mathbf{x}_0^{(n)})}}{\tilde{{\xi}}_{n,h,0}}.
\end{split}	
\end{equation}

\subsection{Resampling}

The resampling scheme $(r_s(i))_{s \in [0:t]}$ is not trivial, indeed choosing resampling schemes that are not looking into future observation vanishes all the effort in building optimal proposals \citep{fearnhead2008computational}. Ideally, resampling should be done according to the smoothing distribution $p(\mathbf{x}_s|\mathbf{y}_{[1:t]},\theta)$ \citep{scharth2016particle}:
\begin{equation}
	p(\mathbf{x}_s|\mathbf{y}_{[1:t]},\theta) = \frac{p(\mathbf{y}_{[s+1:t]}|\mathbf{x}_s,\theta)p(\mathbf{x}_s|\mathbf{y}_{[1:s]},\theta)}{p(\mathbf{y}_{[s+1:t]}|\mathbf{y}_{[1:s]},\theta)} \propto p(\mathbf{y}_{[s+1:t]}|\mathbf{x}_s,\theta)p(\mathbf{x}_s|\mathbf{y}_{[1:s]},\theta),
\end{equation}
which is a combination of the probability of observing the future observations given the current sample $\mathbf{x}_s$ and the filtering distribution. The equivalent of the low-cost case where the proposal distribution approximate $p(\mathbf{x}_s|\mathbf{x}_{s-1}, \mathbf{y}_{[s:s+h]},\theta)$ follows trivially for $t=s+h$.

The quantities involved in the optimal resampling cannot be computed in closed form and they need to be approximated. SMC outputs a particle approximation $(P)^{-1}\sum_{p\in [1:P]} {w}^p_s \delta_{\mathbf{x}_s^p}(\mathbf{x}_s)$ of the filtering distribution $p(\mathbf{x}_s|\mathbf{y}_{[1:s]}, \theta)$ and at the same time algorithm \ref{alg:alternative_proposal_h} gives an approximation $\prod_{n \in [1:N]} \boldsymbol{\tilde{\xi}}_{n,h,s+1}^{(\mathbf{x}_s^{(n)})}$ for $p(\mathbf{y}_{[s+1:t]}|\mathbf{x}_s,\theta)$. It then follows that the approximate optimal resampling is:
\begin{equation}
	r_s(i) \propto {w}^i_s \prod_{n \in [1:N]} \boldsymbol{\tilde{\xi}}_{n,h,s+1}^{((\mathbf{x}_s^i)^{(n)})} \quad \text{for } i \in [1:P],
\end{equation}
with $(\mathbf{x}_s^p)_{p \in [1:P]}$ being the sampled particles at time $s$.


\section{Experiments} \label{sec:experiments}
In this section, we analyse the performance of SMC algorithms when using our approximation of the optimal proposal and resampling scheme. We consider simulated data from two compartmental models: Susceptible-Infected-Susceptible (SIS), Susceptible-Exposed-Infected-Removed (SEIR), which are analysed in Section \ref{subsec:SIS} and Section \ref{subsec:SEIR} respectively. For each model we follow an experimental routine inspired by \cite{ju2021sequential}, comparing our method with the BPF and APF: (i) compare methods based on the effective sample size (ESS) $1 \slash \sum_{i \in [1:P]} (r_s(i))$; (ii) compare methods based on the standard deviation of the estimate of marginal likelihood; (iii) study of the marginal likelihood surface on a grid of parameter values for different $t$ when using our method.

All the experiments are run on 32gb Tesla V100 GPU available on the HEC (High-End Computing) facility from Lancaster University. The code can be found in the GitHub repository ``Optimal\_IBM\_proposal''\\ (https://github.com/LorenzoRimella/Optimal\_IBM\_proposal).

\subsection{Susceptible-infected-susceptible} \label{subsec:SIS}
The SIS model is used in epidemiology to model the spread of a disease in a population when herd immunity is not possible. As already mentioned in Section \ref{subsec:individual_based_models}, we can formulate an individual-based model by having individuals-specific covariates $(\mathbf{w}_n)_{n \in [1:N]}$ and use these covariates to define a unique dynamic per each individual. We have covariates of the form $\mathbf{w}_n=[\mathbf{w}_n^{(1)},\mathbf{w}_n^{(2)}]^\mathrm{T}$ where $\mathbf{w}_n^{(1)}=1$  and $\mathbf{w}_n^{(2)} \sim \mathcal{N}(\bullet|0, 1)$ independently for all $n\in [1:N]$. If not specified otherwise we consider $N=100$, time horizon $t=100$ and data generating parameters (DGP) given by: $\boldsymbol{\beta_0}=[-\log(N-1), 0]^\mathrm{T}$, $\boldsymbol{\beta_\lambda}=[-1, 2]^\mathrm{T}$ , $\boldsymbol{\beta_\gamma}=[-1, -1]^\mathrm{T}$ and $\mathbf{q}_t = \mathbf{q}$ with $\mathbf{q}=[0.8, 0.8]^\mathrm{T}$.

\begin{figure}[httb!]
    \centering
    \includegraphics[width = 0.8\textwidth]{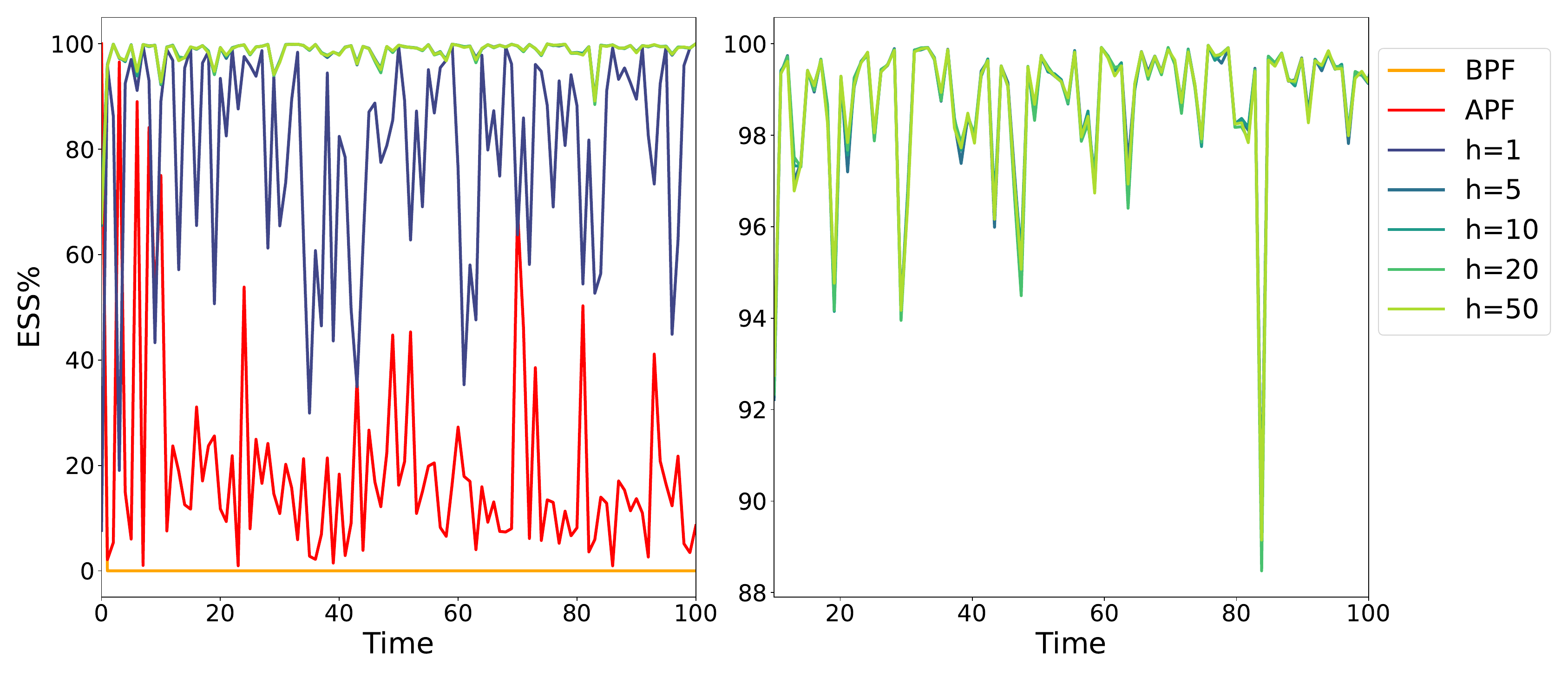}
    \caption{ESS percentage over time for BPF, APF and our method when $h=1,5,10,20,50$. Different colors correspond to different methods. The left plot shows all the listed methods, while the right one considers only $h=10,20,50$ and zoom-in.}
    \label{fig:ESS_SIS}
\end{figure}

The first experiment consists of measuring the ESS for BPF, APF, and $h=1,5,10,20,50$ when $P=512$. Figure \ref{fig:ESS_SIS} displays our findings for a number of particles $P=512$. The BPF fails in sampling any epidemics trajectory, this is due to the mismatch problem mentioned in Section \ref{subsec:inference}, indeed it is enough to mismatch a single individual out of $N$ to assign $0$ probability to the associated particle. APF corrects the proposal by looking at the current observation and so it avoids mismatch. Even though this is a significant improvement compared to BPF the ESS is still very low. Our approximate optimal proposal reaches a significantly better ESS than APF by just looking at the next step in the future ($h=1$). We also observe that choosing $h>5$ does not improve much the performance, this is due to the forgetting property of our HMM \citep{douc2009forgetting}. 

In the next experiment, we look at the standard deviation of the marginal likelihood estimates. We consider two frameworks: one using the data generating parameters and the other substituting $\beta_\lambda$ with $[-3, 0]^\mathrm{T}$. Standard deviations are computed over $100$ runs. The APF is $3-4$ times faster than our method when $h=5$, but the standard deviation is, in both frameworks, $10-20$ times higher than $h=5$ for small $P$ and even $20-30$ times higher than $h=5$ for big $P$. Again, we do not notice a substantial improvement when using $h>5$.
The computational cost highly depends on the implementation, our scripts run on GPUs and parallelize each step of the SMC across individuals and particles, hence we do not report significant changes in the running time when increasing $P$.

\begin{table}[httb!]
    \caption{Table reporting standard deviation for the APF and our method when $h=5,10,20$ under the data generating process (DGP) and non data generating process (NDGP) with $P=128,512,2048$. The mean computational cost of a single step of the SMC is reported in the first row with the name of the algorithm.}
    \label{tab:table_SIS}
    \centering
    \resizebox{12cm}{!}{
    \begin{tabular}{l|cc|cc|cc|cc}
    \hhline{~|--------}
          & APF   & 0.7s   & h=5   & 2.5s   & h=10   & 3.94s   & h=20   & 6.61s   \\
    \hhline{~|--------}
          & DGP   & NDGP   & DGP   & NDGP   & DGP    & NDGP    & DGP    & NDGP    \\
    \hline
     P    & std   & std    & std   & std    & std    & std     & std    & std     \\
     128  & 4.99  & 9.89   & 0.3   & 0.92   & 0.31   & 1.0     & 0.37   & 0.89    \\
     512  & 4.01  & 6.66   & 0.17  & 0.48   & 0.18   & 0.49    & 0.18   & 0.48    \\
     2048 & 2.83  & 6.23   & 0.11  & 0.25   & 0.11   & 0.22    & 0.11   & 0.22    \\
    \hline
    \end{tabular}
    }
\end{table}

\begin{figure}[httb!]
    \centering
    \includegraphics[width = \textwidth]{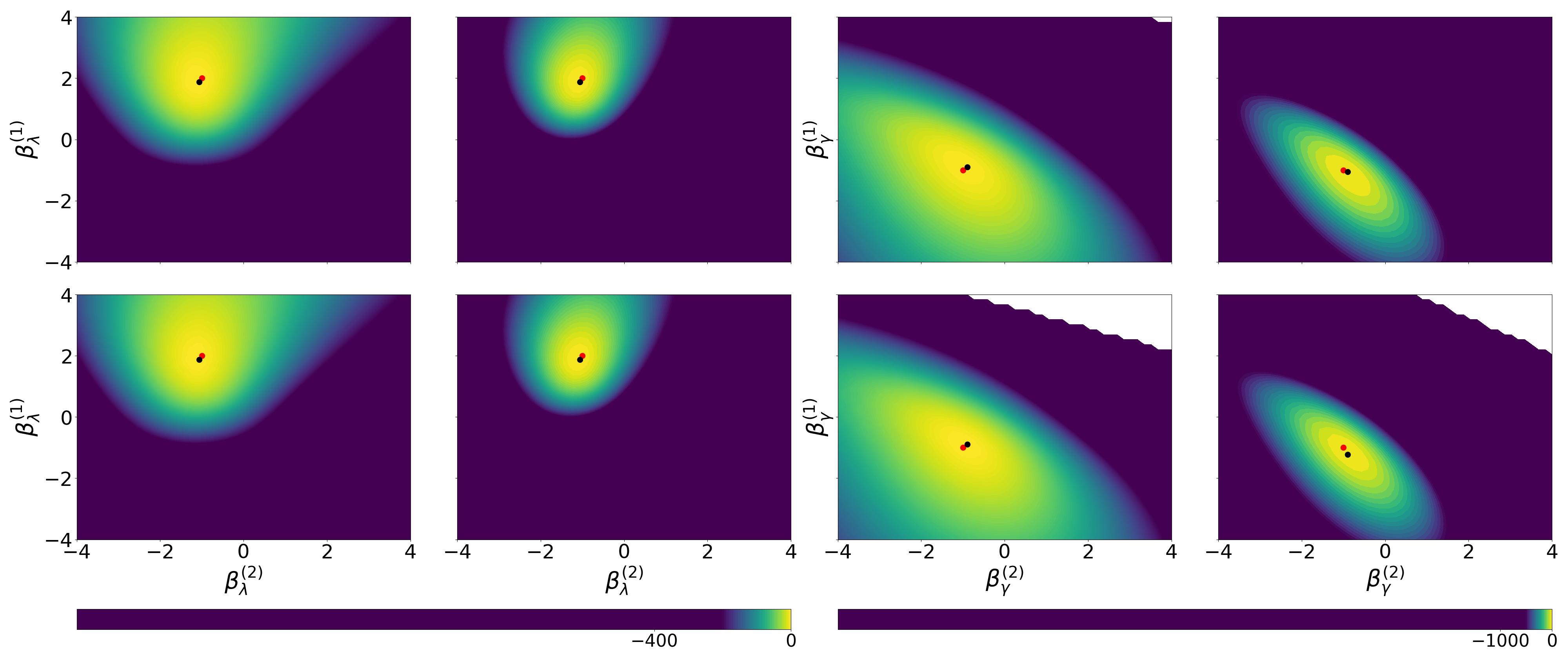}
    \caption{Marginal likelihood contour plots on a $\beta_\lambda$ grid and a $\beta_\gamma$ grid in log-scale. The first and second columns refer to $t=50, 100$ from left to right for $\beta_\lambda$. The third and fourth columns refer to $t=50, 100$ from left to right for $\beta_\gamma$. Rows refer to $h=5, 10$ from top to bottom. The colorbars are common across parameters and their maximum is set to $0$. In red are the data-generating parameters and in black are the MLE on the grid.}
    \label{fig:beta_lambdagamma_SIS}
\end{figure}

Now suppose we want to infer $\beta_\lambda$ or $\beta_\gamma$. We start by setting $\beta_\lambda$ in a 2-dimensional grid on $[-4,4]^2$ and the other parameters to the data generating parameters (including $\beta_\gamma$). We then compute estimates of the marginal likelihood with an SMC employing our proposal and resampling scheme when $P=512$. The procedure is then replicated for $\beta_\gamma$. Both experiments are run for $h=5,10$ and $t=50,100$, with new data generated per each value of $t$. Marginal likelihood contour plots are reported in Figure \ref{fig:beta_lambdagamma_SIS} in log scale and normalized to have their max in zero.

In both figures, we can observe that increasing the time concentrates the likelihood around the data-generating parameters. 
Choosing $h=10$ does not improve much inference over $\beta_\lambda$, but it helps to infer $\beta_\gamma$ by removing some combination of the parameters from the inference (white spaces). 

\subsection{Susceptible-exposed-infected-removed}\label{subsec:SEIR}
The SEIR model is another popular model in epidemiology \citep{he2020seir, deguen2000estimation, porter2013path}, it is used when the disease is expected to have a latent period (exposed compartment) and herd immunity (removed compartment). The SEIR case is significantly more challenging than the SIS because the transition kernel constrains the dynamic on $S \to E \to I \to R$ and so if in our SMC at time $t-1$ we have a particle with individual $n$ in compartment $S$ and we then observe the same individual at time $t$ in compartment $I$ or $R$ the SMC assigns $0$ probability to that particle. 

As for the SIS case, a heterogeneous SEIR model is obtained by including a collection of covariates defining $(\mathbf{w}_n)_{n \in [1:N]}$. The initial distribution $\mathbf{p}_{n,0}$ is defined on compartments $1$ (S) and $3$ (I) as for the SIS case, with zeros for compartments $2$ (E) and $4$ (I). Similarly, $(\mathbf{K}_{n,\bullet})_{n \in [1:N]}$ is defined as the SIS for transitions $1, 2$ (S, E) and $3, 4$ (I, R), with the additional transition $2,3$ (E, I) given by $1-\exp(-\rho)$. Full definitions of $\mathbf{p}_{n,0}$ and $(\mathbf{K}_{n,\bullet})_{n \in [1:N]}$ are available in the appendix. The emission distribution follows \eqref{eq:granular_observation}.

We have covariates of the form $\mathbf{w}_n=[\mathbf{w}_n^{(1)},\mathbf{w}_n^{(2)}]^\mathrm{T}$ where $\mathbf{w}_n^{(1)}=1$  and $\mathbf{w}_n^{(2)} \sim \mathcal{N}(\bullet|0, 1)$ independently for all $n\in [1:N]$. If not specified otherwise we consider $N=1000$, time horizon $t=100$ and data generating parameters given by: $\boldsymbol{\beta_0}=[-\log(N \slash 10 -1), 0]^\mathrm{T}$, $\boldsymbol{\beta_\lambda}=[1, 2]^\mathrm{T}$, $\rho = 0.2$, $\boldsymbol{\beta_\gamma}=[-1, -1]^\mathrm{T}$ and $\mathbf{q}_s = \mathbf{q}$ with $\mathbf{q}=[0, 0, 0.4, 0.6]^\mathrm{T}$.

\begin{figure}[httb!]
    \centering
    \includegraphics[width = 0.8\textwidth]{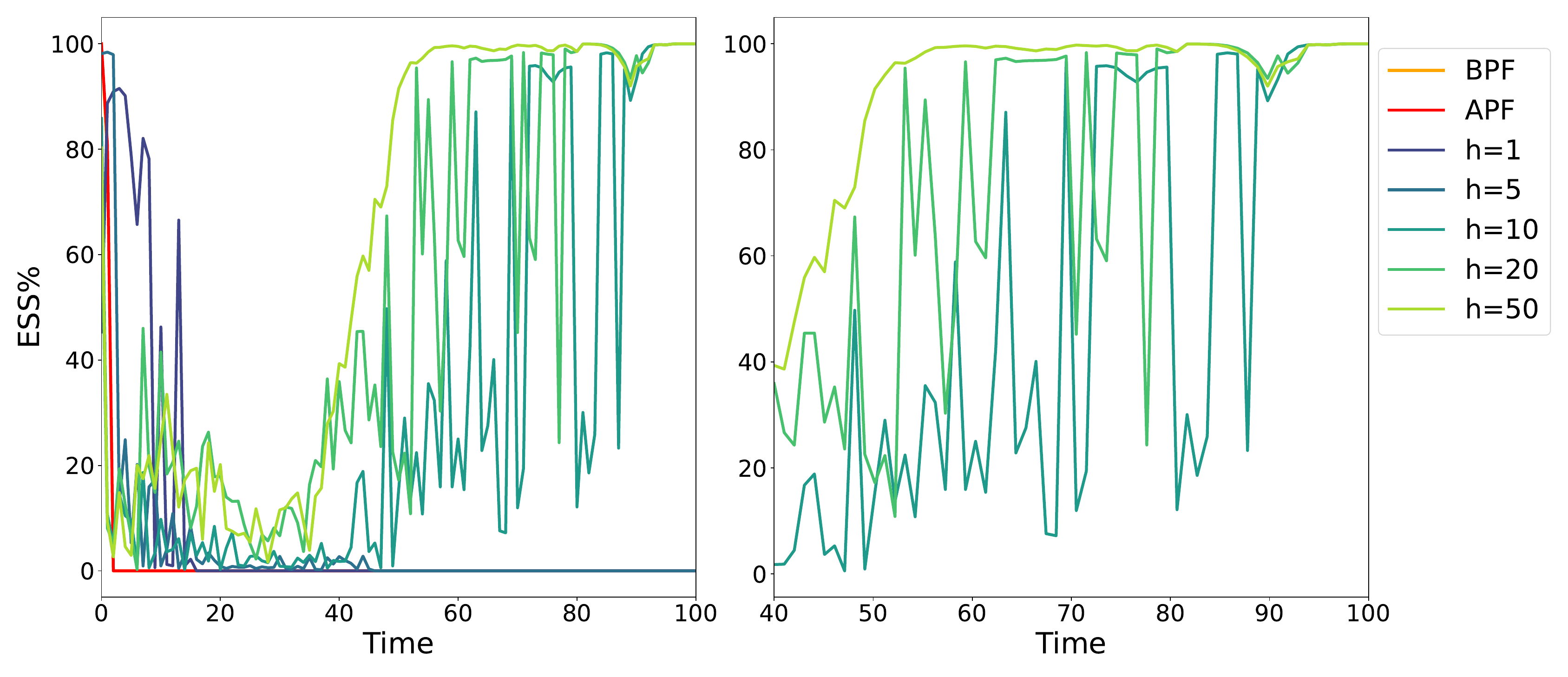}
    \caption{ESS percentage over time for BPF, APF, and our method when $h=1,5,10,20,50$. Different colors correspond to different methods. The left plot shows the listed methods, while the right one considers only $h=5,10,20,50$ and zoom-in.}
    \label{fig:ESS_SEIR}
\end{figure}

As for the SIS case, we start by analysing the ESS for BPF, APF, and $h=1,5,10,20,50$ with $P=512$. In Figure \ref{fig:ESS_SEIR} both the BPF and the APF fail due to a mismatch between the proposed particles and the observations. Even our method fails for $h\leq 5$, but when choosing $h\geq 10$ we are able to avoid mismatch and get an increasing in time ESS. 

\begin{table}[httb!]
    \caption{Table reporting standard deviation for our method when $h=5,10,20, 50$ under the data generating parameters (DGP) and non data generating parameters (NDGP) with $P=128,512,2048$. The mean computational cost of a single step of the SMC is reported in the first row with the name of the algorithm.}
    \label{tab:table_SEIR}
    \centering
    \resizebox{12cm}{!}{
    \begin{tabular}{l|ll|ll|ll|ll}
    \hhline{~|--|--|--|--}
          & h=5   & 0.9s   & h=10   & 3.5s   & h=20   & 5.45s   & h=50   & 9.03s   \\
    \hhline{~|--|--|--|--}
          & DGP   & NDGP   & DGP    & NDGP   & DGP    & NDGP    & DGP    & NDGP    \\
    \hline
     P    & std   & std    & std    & std    & std    & std     & std    & std     \\
     128  & 58.18 & 68.2   & 20.47  & 32.6   & 9.59   & 18.32   & 6.93   & 11.71   \\
     512  & 48.23 & 74.78  & 18.37  & 28.64  & 6.39   & 15.76   & 6.23   & 10.72   \\
     2048 & 42.7  & 58.37  & 15.03  & 24.68  & 5.69   & 13.25   & 4.57   & 10.45   \\
    \hline
    \end{tabular}
    }
\end{table}

We then investigate the standard deviation and computational cost of our method when $h$ changes, and we report our results in Table \ref{tab:table_SEIR}. Observe there is a significant improvement in the standard deviation when increasing $h$ up to $50$, with the jump from $h=20$ to $h=50$ being less substantial. Clearly, there is a trade-off, a decrease in standard deviation has to be paid for by an increase in computational cost, but it seems worth it for $h < 50$, given that halving the standard deviation is associated with less than doubling the computational cost.

\begin{figure}[httb!]
    \centering
    \includegraphics[width = \textwidth]{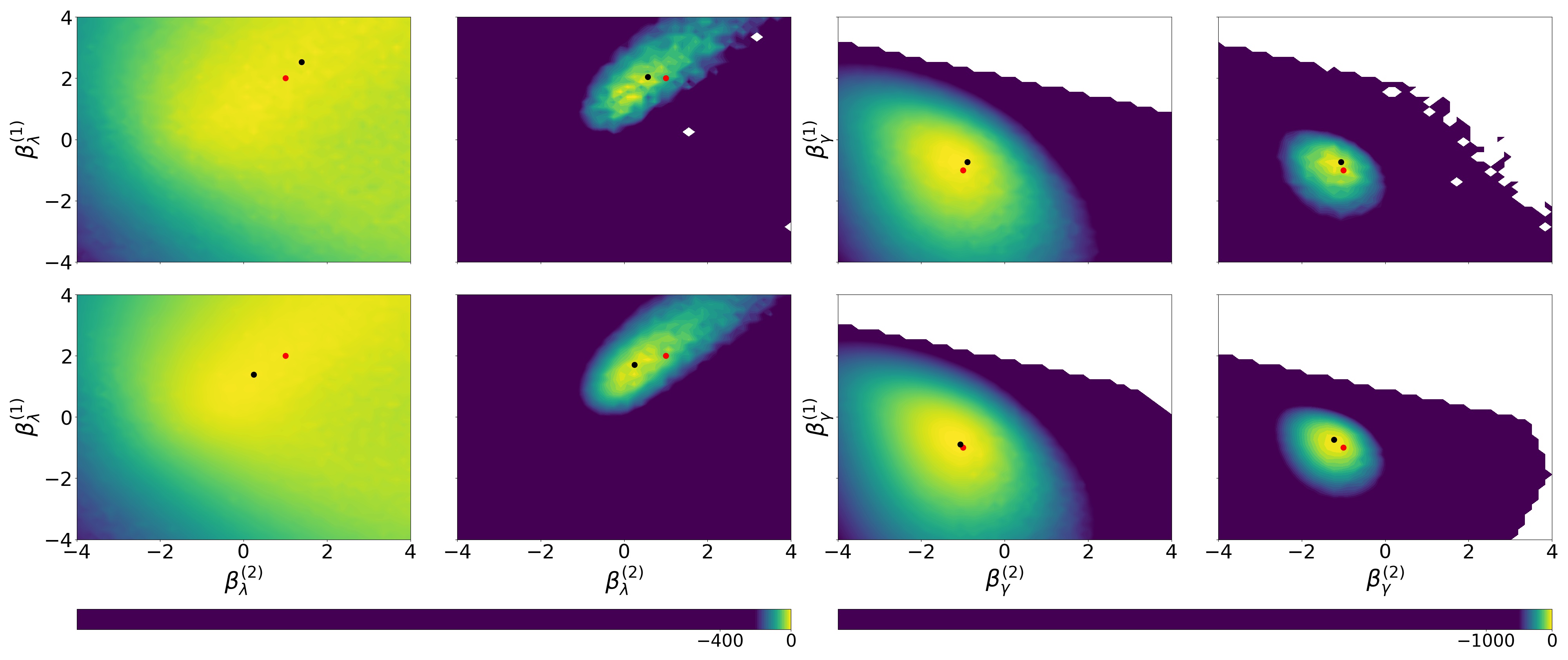}
    \caption{Marginal likelihood contour plots on a $\beta_\lambda$ grid and a $\beta_\gamma$ grid in log-scale. The first and second columns refer to $t=50, 100$ from left to right for $\beta_\lambda$. The third and fourth columns refer to $t=50, 100$ from left to right for $\beta_\gamma$. Rows refer to $h=5, 10$ from top to bottom. The colorbars are common across parameters and their maximum is set to $0$. In red are the data-generating parameters and in black are the MLE on the grid.}
    \label{fig:beta_lambdagamma_SEIR}
\end{figure}

We conclude by reproducing the marginal likelihood surfaces of $\beta_\lambda$ and $\beta_\gamma$ on grids for the SEIR scenario. The experiments are run for $h=10, 20$ and $t=50,100$, with new data generated per each value of $t$. Figure \ref{fig:beta_lambdagamma_SEIR} shows the marginal likelihood contour plots on a log scale and normalized to have their max in zero. As for the SIS case, an increase in $t$ concentrates the likelihood around the DGP as can be seen in both figures. Notice that the log-likelihood surface of $\beta_\lambda$ is multi-modal, this is due to observing neither susceptible nor exposed individuals, which makes inference on this parameter significantly harder. Choosing $h=20$ seems to smooth the likelihood surface and it also avoids failure close to the data generating parameters, as seen by white holes in the surface for $h=10$ and $t=100$. $\beta_\gamma$ has a smoother surface compare to $\beta_\lambda$ and again increasing $h$ seems to improve the shape.

\section{Discussion} \label{sec:discussion}

Our findings demonstrate the difficulties in fitting individual-based epidemic models in the presence of censored data and highlight the significance of incorporating future observations to guide the choice of proposal distributions in SMC algorithms. The underlying framework in which proposal distributions are built is general and the algorithm requires only obtaining, for each individual, estimates of the transition rates at the times $[t+1,t+h]$, which are then propagated backwards to build a proposal distribution that includes future observations.

While the overall procedure has been presented as an algorithm to compute good proposal distributions, it seems like several aspects of the work could be used to improve existing methods. For example, as already mentioned, our backward recursion method could be used to compute the marginal likelihood approximation in pseudo-likelihood methods \citep{andrieu2009pseudo}, or as the first approximate model stage in the delayed acceptance scheme of \citet{golightly2015delayed}.

Our implementation focuses on the case of homogeneous reporting rates in a fully connected population, however, the work can be extended beyond this setting. Indeed, it is straightforward to use these techniques for heterogeneous reporting rates, as discussed at the end of Section \ref{sec:whiteley_rime_app}, and for spatial epidemic models we would simply need to obtain an estimate of the spatial risk of infection to be able to run the recursion. Epidemic models with an open population (e.g. migration or births-deaths) and misreporting can also be included in the class of models we can deal with, by substituting the multinomial approximation \citep{whiteley2021inference} with alternative approximations \citep{whitehouse2022consistent}.

\section*{Supplementary Materials}

The supplementary materials are divided in five sections reporting:
\begin{enumerate}
    \item the main notation and conventions;
    \item an introduction to compartmental model;
    \item the main computation in HMM and SMC;
    \item the algorithm by \cite{whiteley2021inference};
    \item some additional experiments and extra details on some experiments from the main paper.
\end{enumerate}
\par

\section*{Acknowledgements}

This work is supported by EPSRC grants EP/R018561/1 (Bayes4Health) and EP/R034710/1 (CoSInES). The authors thank Simon Spencer for a helpful discussion that initiated the idea. They also thank Kunyang He, Yize Hao, Edward L. Ionides, Nick Whiteley, and Michael Whitehouse for identifying a misalignment between the SMC algorithm and the likelihood estimate in the paper and the SMC algorithm and the likelihood estimate in the code. These issues have been corrected in this version of the manuscript. Both the algorithm and the likelihood estimate now align with the code in the GitHub repository: https://github.com/LorenzoRimella/Optimal-IBM-proposal.

\par


\bibhang=1.7pc
\bibsep=2pt
\fontsize{9}{14pt plus.8pt minus .6pt}\selectfont
\renewcommand\bibname{\large \bf References}
\expandafter\ifx\csname
natexlab\endcsname\relax\def\natexlab#1{#1}\fi
\expandafter\ifx\csname url\endcsname\relax
  \def\url#1{\texttt{#1}}\fi
\expandafter\ifx\csname urlprefix\endcsname\relax\def\urlprefix{URL}\fi

\bibliographystyle{chicago}      
\bibliography{references}   

\appendix

\section{Preliminaries} \label{sec:notation_app}

Given the vector $\mathbf{a}$ we use $\mathbf{a}^{(i)}$ for the $(i)$-th element of $\mathbf{a}$ (vectors are always assumed to be column vectors), given the matrix $\mathbf{A}$ we use $\mathbf{A}^{(i,j)}$ for the $(i,j)$-th element of $\mathbf{A}$ and we use $\mathbf{A}^{(i,\bullet)}$ (or $\mathbf{A}^{(\bullet,j)}$) to represent the column vector given by the $(i)$-th column (or the $(j)$-th row) of matrix $\mathbf{A}$. We use $\mathbf{A}^{\mathrm{T}}$ for the transpose of matrix $\mathbf{A}$ (similarly for vectors). The dot product between two matrices $\mathbf{A},\mathbf{B}$ is denoted by $\mathbf{A}\mathbf{B}$ (similarly for vectors). Sums or differences between vectors or matrices have to be thought of as elementwise, similarly when using vectors or matrices and a scalar. We employ the convention $\mathbb{R}_+$ for the positive real numbers. We use the notation $\mathbb{I}_{a}(b)$ for the indicator function, with $a,b$ being scalars, vectors, or matrices.

Given the probability mass (or density) function of a random variable $\mathcal{D}(x|\theta)$ evaluated in $x$ with parameters $\theta$ as in table \ref{tab:notation}, we use $X \sim \mathcal{D}(\bullet|\theta)$ to say the random variable $X$ is drawn from the corresponding probability random variable and we denote with $\mathbb{E}_{\mathcal{D}(X|\theta)}[X]$ the corresponding expectation.

\begin{table}[httb!]
    \centering
    \resizebox{14cm}{!}{
    \begin{tabular}{l|llllll}
    \hline
    Distribution & Categorical          & Bernoulli           & Binomial & Gaussian & Uniform & Multinomial \\
    \hline
    Notation & $\mathcal{C}at_M(i|\mathbf{p})$ & $\mathcal{B}e(i|q)$ & $\mathcal{B}in(i|N,q)$ & $\mathcal{N}(a|\mu, \sigma^2)$ & $\mathcal{U}nif(q|a,b)$ & $\mathcal{M}ult(\mathbf{c}|N, \mathbf{p})$\\
    \hline
    Domain & $[1:M]$ & $\{0,1\}$ & $[0:N]$ & $\mathbb{R}$ & $[a,b]$ & $\{\mathbf{c}: \sum_{i \in [1:M]} \mathbf{c}^{(i)}=N\}$\\
    \hline
    \end{tabular}
    }
    \caption{Notation table for probability mass and density functions along with domains.}
    \label{tab:notation}
\end{table}

\section{Introduction to compartmental models} \label{sec:comp_model_app}
In epidemiology, compartmental models are used to describe the dynamics of an epidemic in a population, where the compartments represent different stages of the disease \citep{brauer2008compartmental}. A closed population stochastic compartmental model is fully defined by: the number of compartments $M$, the population size $N$, the initial probability of being assigned to a compartment $\mathbf{p}_0$, and the probability of transition from one compartment to the other $\mathbf{K}_{\bullet}$, which is a stochastic transition matrix $\mathbf{c} \to \mathbf{K}_{\mathbf{c}}$ defined as a function of the compartments' state $\mathbf{c}$. The dynamics of a compartmental model is generally described by a discrete-time Markov chain $(\mathbf{c}_t)_{t \geq 0}$ counting the number of individuals in each compartment, i.e. $\mathbf{c}_t$ is an $M$-dimensional vector with $\mathbf{c}_t^{(i)}$ being the number of individuals in compartment $i$ \citep{keeling2011modeling}. The evolution of $\mathbf{c}_t$ can be specified from an individual perspective by defining the discrete-time Markov chain $(\mathbf{x}_t)_{t \geq 0}$, where $\mathbf{x}_t$ is an $N$-dimensional vector representing the state of the population at time $t$ \citep{whiteley2021inference}, i.e. $\mathbf{x}_t^{(n)}$ is the compartment of individual $n$ at time $t$. The evolution of $(\mathbf{x}_t)_{t \geq 0}$ and $(\mathbf{c}_t)_{t \geq 0}$ can be then represented as the following flow:
\begin{itemize}
	\item[Time $0$:] $\mathbf{x}^{(n)}_0 \sim \mathcal{C}at_{M}(\bullet| \mathbf{p}_0)$ for $n \in [1:N]$ and\\ $\mathbf{c}_0^{(i)} = \sum_{n=1}^N \mathbb{I}_{\mathbf{x}_0^{(n)}}(i)$ for $i \in [1:M]$;
	\item[Time $t$:] $\mathbf{x}^{(n)}_t|\mathbf{x}_{t-1} \sim \mathcal{C}at_{M} \left ( \bullet \Big{|}  \mathbf{K}_{\mathbf{c}_{t-1}}^{(\mathbf{x}_{t-1}^{(n)}, \bullet)} \right )$ for $n \in [1:N]$ and\\ $\mathbf{c}_t^{(i)} = \sum_{n=1}^N \mathbb{I}_{\mathbf{x}_t^{(n)}}(i)$ for $i \in [1:M]$;
\end{itemize}             
where we first make all the individuals move (simulate $\mathbf{x}_t$) and then count the individuals in each compartment (compute $\mathbf{c}_t$).

\paragraph{SIS example}
The susceptible-infected-susceptible model (SIS) is a well-known compartmental model used to model the spread of a disease in a population where herd immunity is not possible, i.e. the individuals can be re-infected. A stochastic SIS with closed population $N$ can be represented in the previous framework by: $M=2$, $\mathbf{p}_0$ in the $2$-dimensional simplex (with $\mathbf{p}_0^{(1)}$ probability of being susceptible at time $0$) and $\mathbf{K}_{\bullet}$ a $2\text{by}2$-dimensional stochastic transition matrix. A popular choice of $\mathbf{K}_{\bullet}$ is:
$$
\mathbf{K}_{c} = 
\begin{bmatrix}
	  e^{-\beta \frac{\mathbf{c}^{(2)}}{N}} & 1-e^{-\beta \frac{\mathbf{c}^{(2)}}{N}}\\
	1-e^{-\gamma} &   e^{-\gamma}
\end{bmatrix}
$$    
with $\beta, \gamma \in \mathbb{R}_+$ transmission and recovery parameters.

Focusing on the compartments' state automatically assumes homogeneous individuals, which is a significant simplification of the real world, where each individual often has their own covariates, e.g. age. individual-based compartmental models or simply individual-based models relax the homogeneity assumption and look at the disease from an individual perspective. The main difference resides in the individual-specific $(\mathbf{p}_{n,0})_{n \in [1:N]}$ and $(\mathbf{K}_{n,\bullet})_{n \in [1:N]}$ representing the heterogeneous dynamic of the individuals. 

\subsection{Considerations on non-granular observation models}
Compartmental models in epidemiology are generally treated as latent and accompanied by an observation model representing the conditional distribution of the observations given the current compartments' state often refer as the emission distribution. Popular choices are the Negative binomial distribution \citep{fintzi2021linear} and the binomial distribution \citep{lekone2006statistical}, with the latter being used by \cite{ju2021sequential} in an individual-based model framework. The main limitation of these emission distributions is that they are formulated over the aggregated population, while for an individual-based model, it is fair to assume individual-based observations. 

\section{Hidden Markov models} \label{sec:HMM_app}
In an HMM $(\mathbf{x}_t, \mathbf{y}_t)_{t \geq 1}$ the recursive computation of $p(\mathbf{x}_t|\mathbf{y}_{[1:t]},\theta)$ and $p(\mathbf{y}_{[1:t]}|\theta)$ is known as forward algorithm and it consists of the following steps:
\begin{itemize}
	\item[Time $0$:] $p(\mathbf{x}_0|\theta)$ from the initial distribution;
	\item[Time $t$:] $p(\mathbf{x}_t|\mathbf{y}_{[1:t-1]},\theta) = \sum_{\mathbf{x}_{t-1} \in [1:M]^N} p(\mathbf{x}_t| \mathbf{x}_{t-1},\theta) p(\mathbf{x}_{t-1}|\mathbf{y}_{[1:t-1]},\theta)$ and \\
	$p(\mathbf{y}_t|\mathbf{y}_{[1:t-1]},\theta) = \sum_{\mathbf{x}_{t} \in [1:M]^N} p(\mathbf{y}_t| \mathbf{x}_{t},\theta) p(\mathbf{x}_t|\mathbf{y}_{[1:t-1]},\theta)$ and \\
	$p(\mathbf{x}_t|\mathbf{y}_{[1:t]},\theta) = \frac{p(\mathbf{y}_t| \mathbf{x}_{t},\theta) p(\mathbf{x}_t|\mathbf{y}_{[1:t-1]},\theta)}{p(\mathbf{y}_t|\mathbf{y}_{[1:t-1]},\theta)}$ and \\
	$p(\mathbf{y}_{[1:t]}|\theta) = p(\mathbf{y}_{[1:t-1]}|\theta)p(\mathbf{y}_t|\mathbf{y}_{[1:t-1]},\theta)$.
\end{itemize}
A close-form solution for the filtering is often not available and an SMC algorithm can be employed to compute particle estimates of both $p(\mathbf{x}_t|\mathbf{y}_{[1:t]},\theta)$ and $p(\mathbf{y}_{[1:t]}|\theta)$. A general SMC algorithm is presented in \ref{alg:SMC_general} and it consists of: resampling the previous particles according to the resampling scheme and correcting the weights accordingly, proposing new particles with the proposal distribution, update the weights. The likelihood estimate is then given by $\hat{p}(y_{[1:s]}|\theta) \leftarrow \hat{p}(y_{[1:s-1]}|\theta) (P)^{-1} \sum_{p \in [1:P]} {w}_s^p$, where ${w}_s^p$ are the weights after update.

\begin{algorithm}
	\caption{A general sequential Monte Carlo algorithm}\label{alg:SMC_general}
	\begin{algorithmic}[1]
		\Require{$P$, $\theta$, $\mathbf{y}_{[1:t]}$, $(q(\mathbf{x}_s|\mathbf{x}_{s-1}, \mathbf{y}_{[1:t]}))_{s \in [0:t]}$, $(r_s(i))_{s \in [0:t]}$}
		\State Sample $\mathbf{x}_0^p \sim q(\bullet|\mathbf{y}_{[1:t]})$
		\State Compute $w_0^p \leftarrow \frac{p(\mathbf{x}_0^p|\theta)} {q(\mathbf{x}_0^p|\mathbf{y}_{[1:t]})}$ for $p \in [1:P]$ 
		\For{$s = 1,\dots, t$}
                \State Set $\bar{w}_{t-1}^p \propto {w}_{t-1}^p$ and normalize for $p \in [1:P]$ 
  			\State Resample $i_{s-1}^p \sim r_{s-1}(\bullet)$ and set $\tilde{\mathbf{x}}_{s-1}^p \leftarrow \mathbf{x}_{s-1}^{i_{s-1}^p}$ for $p \in [1:P]$ 
                \State Correct the weights $\tilde{w}_{s-1}^p = \frac{\bar{w}_{s-1}^{i_{s-1}^p}}{r_{s-1}(i_{s-1}^p)}$ for $p \in [1:P]$
			\State Propose $\mathbf{x}_s^p \sim q(\bullet|\tilde{\mathbf{x}}_{s-1}, \mathbf{y}_{[1:t]})$
			\State Compute $w_s^p \leftarrow \tilde{w}^p_{s-1} \frac{p(\mathbf{x}_s^p|\tilde{\mathbf{x}}_{s-1}^p,\theta)p(\mathbf{y}_s|\mathbf{x}_{s}^p,\theta)}{q(\mathbf{x}_s^p|\tilde{\mathbf{x}}_{s-1}^p, \mathbf{y}_{[1:t]})}$ for $p \in [1:P]$ 
		\EndFor
		\State $\hat{p}(y_{[1:t]}|\theta) \leftarrow \prod_{s \in[1:t]} (P)^{-1} \sum_{p \in [1:P]} {w}_s^p$
	\end{algorithmic}
\end{algorithm}

In the HMM terminology the initial distribution $p(\mathbf{x}_0|\theta)$, the transition kernel $p(\mathbf{x}_t|\mathbf{x}_{t-1},\theta)$ and the emission distribution $p(\mathbf{y}_t|\mathbf{x}_t,\theta)$ for our individual-based model with granular observation are given by:
\begin{equation} \label{eq:HMM}
	\begin{split}
		&p(\mathbf{x}_0|\theta) = \prod_{n \in [1:N]} \mathbf{p}_{n,0}^{(\mathbf{x}_0^{(n)})}, \quad p(\mathbf{x}_t|\mathbf{x}_{t-1},\theta) = \prod_{n \in [1:N]} \mathbf{K}_{n, \mathbf{c}_{t-1}}^{(\mathbf{x}_{t-1}^{(n)},\mathbf{x}_t^{(n)})}, \\
		&p(\mathbf{y}_t|\mathbf{x}_t,\theta)= \prod_{n \in [1:N]}  \left ( \mathbf{q}_t^{(\mathbf{x}_t^{(n)})} \right )^{\mathbb{I}_{\mathbf{y}_t^{(n)}}(\mathbf{x}_t^{(n)})} \left ( 1- \mathbf{q}_t^{(\mathbf{x}_t^{(n)})} \right )^{\mathbb{I}_{\mathbf{y}_t^{(n)}}(0)}.
	\end{split}
\end{equation}
These definitions are useful to see how the transition kernel and the emission distribution can be used to simplify the formulation of $p(\mathbf{y}_t|\mathbf{x}_{t-1},\theta)$:
\begin{equation}\label{eq:factorization_initial_step}
\begin{split}
	p(\mathbf{y}_t|\mathbf{x}_{t-1},\theta) &= \sum_{\mathbf{x}_t \in [1:M]^N} \prod_{n \in [1:N]} \mathbf{K}_{n, \mathbf{c}_{t-1}}^{(\mathbf{x}_{t-1}^{(n)},\mathbf{x}_t^{(n)})} \left ( \mathbf{q}_t^{(\mathbf{x}_t^{(n)})} \right )^{\mathbb{I}_{\mathbf{y}_t^{(n)}}(\mathbf{x}_t^{(n)})} \left ( 1- \mathbf{q}_t^{(\mathbf{x}_t^{(n)})} \right )^{\mathbb{I}_{\mathbf{y}_t^{(n)}}(0)}\\
	&= 
	\prod_{n \in [1:N]} \sum_{\mathbf{x}^{(n)}_t \in [1:M]} \mathbf{K}_{n, \mathbf{c}_{t-1}}^{(\mathbf{x}_{t-1}^{(n)},\mathbf{x}_t^{(n)})} \left ( \mathbf{q}_t^{(\mathbf{x}_t^{(n)})} \right )^{\mathbb{I}_{\mathbf{y}_t^{(n)}}(\mathbf{x}_t^{(n)})} \left ( 1- \mathbf{q}_t^{(\mathbf{x}_t^{(n)})} \right )^{\mathbb{I}_{\mathbf{y}_t^{(n)}}(0)}.
\end{split}
\end{equation} 

\begin{algorithm}
	\caption{Multinomial approximation by \cite{whiteley2021inference}}\label{alg:multinomial_approx}
	\begin{algorithmic}[1]
		\Require{$(\mathbf{\bar{p}}_{n,0})_{n \in [1:N]})$, $(\mathbf{\bar{K}}_{\bullet})_{n \in [1:N]}$, $(\mathbf{q}_s)_{s \in [t]}$, $(\mathbf{o}_s)_{s \in [t]}$}
		\State $\mathbf{m}_{0|0} \leftarrow \mathbf{\bar{p}}_{n,0}$  \Comment{Forward step}
		\For{$s = 1,\dots, t$}
		\State $\mathbf{m}_{s-1|s} \leftarrow \left ( \mathbf{m}_{s-1|s-1}^{\mathrm{T}} \mathbf{\bar{K}}_{\mathbf{m}_{s-1}} \right )^{\mathrm{T}}$
		\State $\mathbf{m}_{s|s} \leftarrow \frac{\mathbf{o}_s}{N} + \left (  1 - \frac{\mathbf{1}_M^{\mathrm{T}}\mathbf{o}_s}{N}\right ) \frac{\mathbf{m}_{s-1|s} \circ (\mathbf{1}_M-\mathbf{q}_s)}{1-\mathbf{m}_{s-1|s}^{\mathrm{T} }\mathbf{q}_s}$ 
		\EndFor
		\For{$s = t-1,\dots,0$} \Comment{Backward step}
		\State $\mathbf{L}_s \leftarrow \left \{ \left [ (\mathbf{m}_{s|t} \mathbf{1}_M^{\mathrm{T}}) \circ \mathbf{\bar{K}}_{\mathbf{m}_s} \right] \slash \left [ \mathbf{1}_M (\mathbf{m}_{s|t}^{\mathrm{T}} \mathbf{\bar{K}}_{\mathbf{m}_{s}}) \right ] \right \}^{\mathrm{T}}$
		\State $\mathbf{m}_{s|t} \leftarrow \left ( \mathbf{m}_{s+1|T}^{\mathrm{T}} \mathbf{L}_s \right )^{\mathrm{T}}$
		\EndFor
	\end{algorithmic}
\end{algorithm} 

\section{A priori estimate of the compartments' state} \label{sec:whiteley_rime_app}
The approximate homogeneous dynamic considered in the main paper is:
\begin{itemize}
	\item[Time $0$:] $\mathbf{x}^{(n)}_0 \sim \mathcal{C}at_{M}(\bullet| \mathbf{\bar{p}}_{n,0})$ for $n \in [1:N]$ and\\ $\mathbf{c}_0^{(i)} = \sum_{n=1}^N \mathbb{I}_{\mathbf{x}_0^{(n)}}(i)$ for $i \in [1:M]$;
	\item[Time $t$:] $\mathbf{x}^{(n)}_t|\mathbf{x}_{t-1} \sim \mathcal{C}at_{M} \left ( \bullet \Big{|}  \mathbf{\bar{K}}_{\mathbf{c}_{t-1}}^{(\mathbf{x}_{t-1}^{(n)}, \bullet)} \right )$ for $n \in [1:N]$ and\\ $\mathbf{c}_t^{(i)} = \sum_{n=1}^N \mathbb{I}_{\mathbf{x}_t^{(n)}}(i)$ for $i \in [1:M]$.
\end{itemize} 

The full algorithm by \citep{whiteley2021inference} is reported in algorithm \ref{alg:multinomial_approx}.

\section{Additional experiments} \label{sec:experiments_app}

\subsection{Susceptible-infected-susceptible}

\begin{figure}[httb!]
    \centering
    \includegraphics[width =\textwidth]{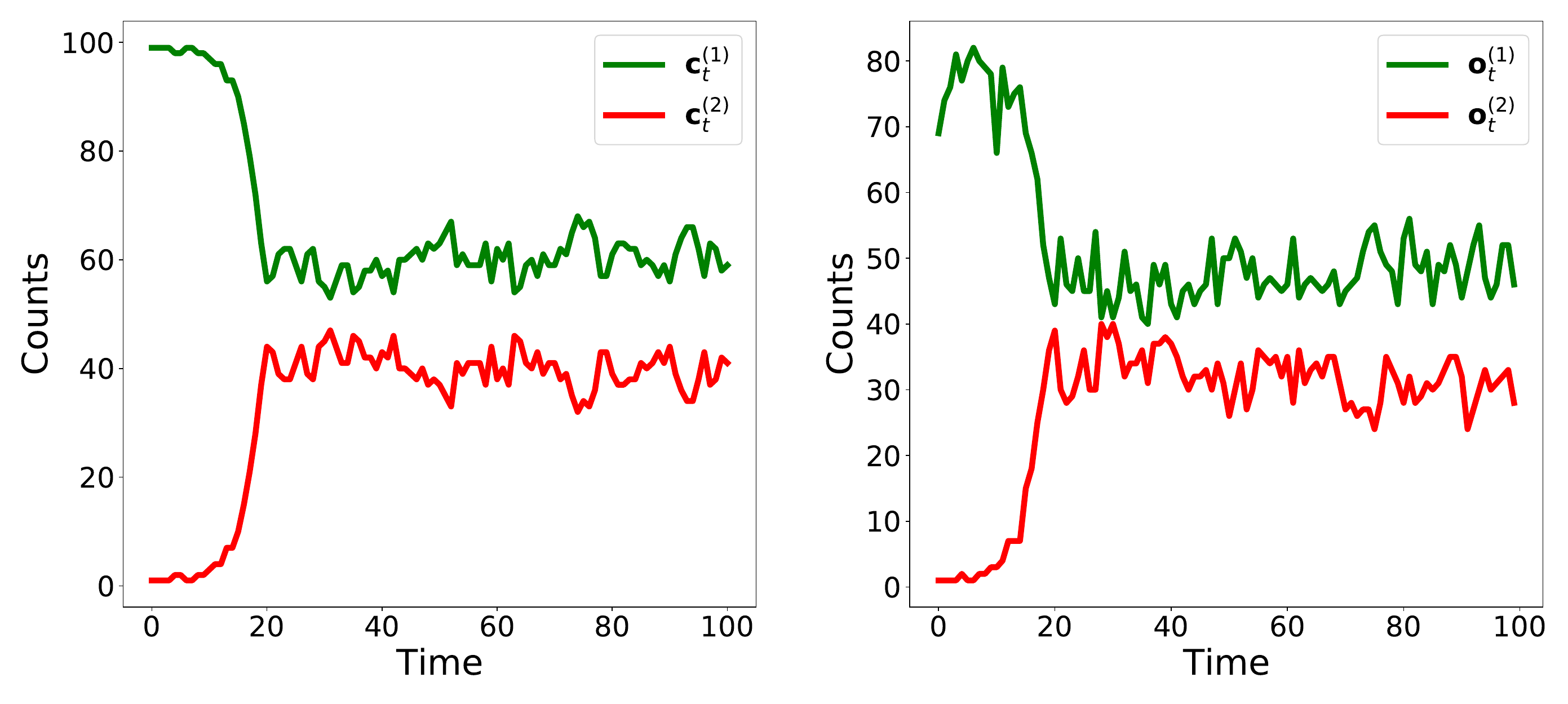}
    \caption{SIS epidemic, on the left the unobserved process, on the right the observed data.}
    \label{fig:SIS_data}
\end{figure}

\begin{table}[httb!]
    \centering
    \resizebox{11cm}{!}{
    \begin{tabular}{l|cc|cc|cc|cc}
    \hhline{~|--------}
          & APF   & 0.7s   & h=5   & 2.5s   & h=10   & 3.94s   & h=20   & 6.61s   \\
    \hhline{~|--------}
          & DGP   & NDGP   & DGP   & NDGP   & DGP    & NDGP    & DGP    & NDGP    \\
    \hline
     P    & std   & std    & std   & std    & std    & std     & std    & std     \\
     64   & 7.02  & 10.97  & 0.4   & 1.26   & 0.48   & 1.35    & 0.46   & 1.19    \\
     128  & 4.99  & 9.89   & 0.3   & 0.92   & 0.31   & 1.0     & 0.37   & 0.89    \\
     256  & 5.24  & 8.3    & 0.27  & 0.72   & 0.24   & 0.67    & 0.35   & 0.63    \\
     512  & 4.01  & 6.66   & 0.17  & 0.48   & 0.18   & 0.49    & 0.18   & 0.48    \\
     1024 & 3.42  & 7.17   & 0.15  & 0.34   & 0.21   & 0.33    & 0.15   & 0.35    \\
     2048 & 2.83  & 6.23   & 0.11  & 0.25   & 0.11   & 0.22    & 0.11   & 0.22    \\
    \hline
    \end{tabular}
    }
    \caption{Table reporting standard deviation for APF and our method when $h=5,10,20$ under DGP and NDGP with $P=64,128,256,512,1024,2048$. The mean computational cost is reported in the first row with the name of the algorithm.}
    \label{tab:table_SIS_full}
\end{table}

\begin{figure}[httb!]
    \centering
    \includegraphics[width = \textwidth]{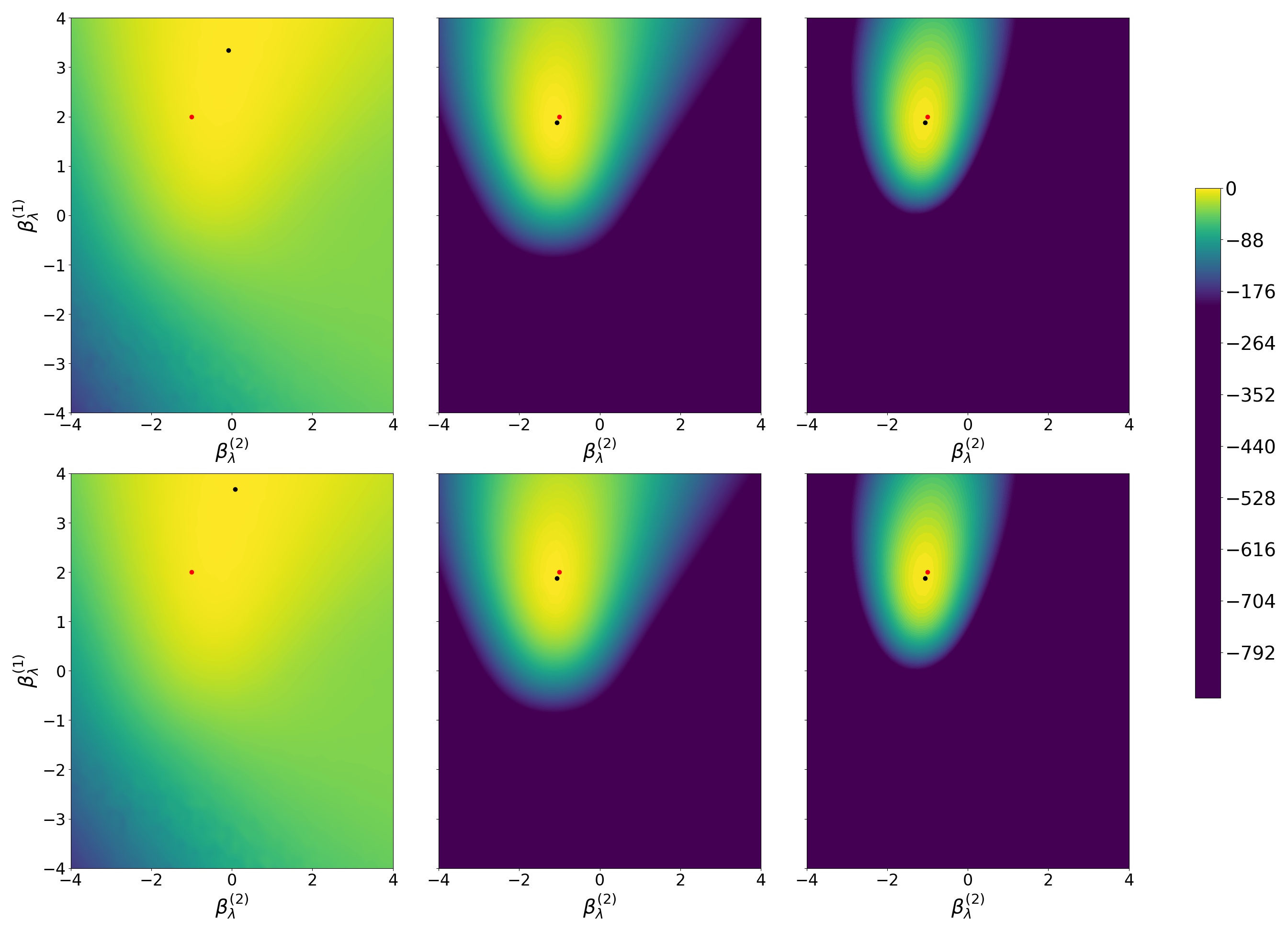}
    \caption{Marginal likelihood contour plots on a $\beta_\lambda$ grid in log-scale. Columns refer to $t=10, 50, 100$ from left to right. Rows refer to $h=5, 10$ from top to bottom. The colorbar is common across the plots and in each plot, the maximum is set to $0$. In red is the DGP and in black is the MLE on the grid.}
    \label{fig:beta_lambda_SIS}
\end{figure}

\begin{figure}[httb!]
    \centering
    \includegraphics[width = \textwidth]{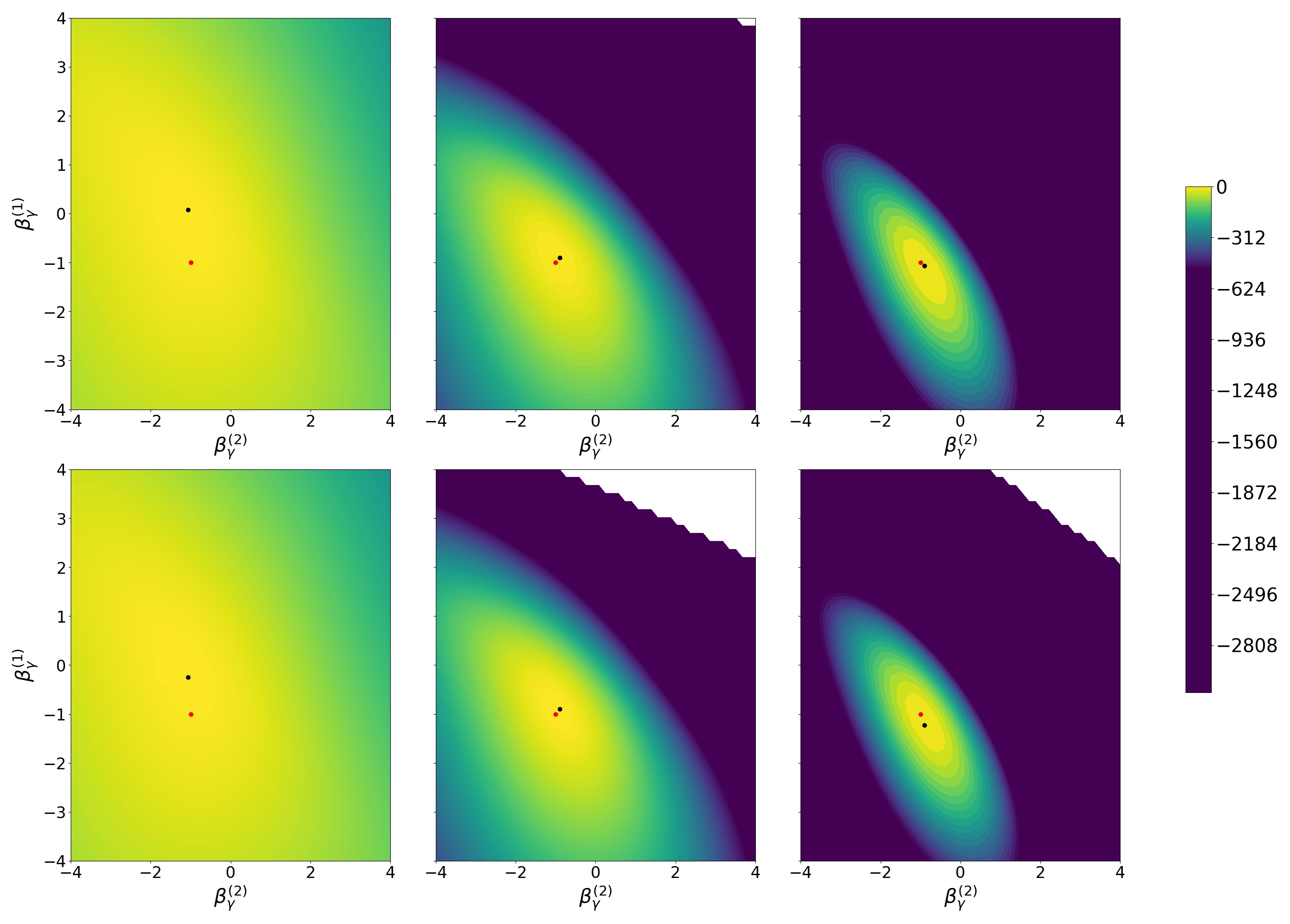}
    \caption{Marginal likelihood contour plots on a $\beta_\gamma$ grid in log-scale. Columns refer to $t=10, 50, 100$ from left to right. Rows refer to $h=5, 10$ from top to bottom. The colorbar is common across the plots and in each plot, the maximum is set to $0$. In red is the DGP and in black is the MLE on the grid.}
    \label{fig:beta_gamma_SIS}
\end{figure}

\begin{figure}[httb!]
    \centering
    \includegraphics[width = 0.8\textwidth]{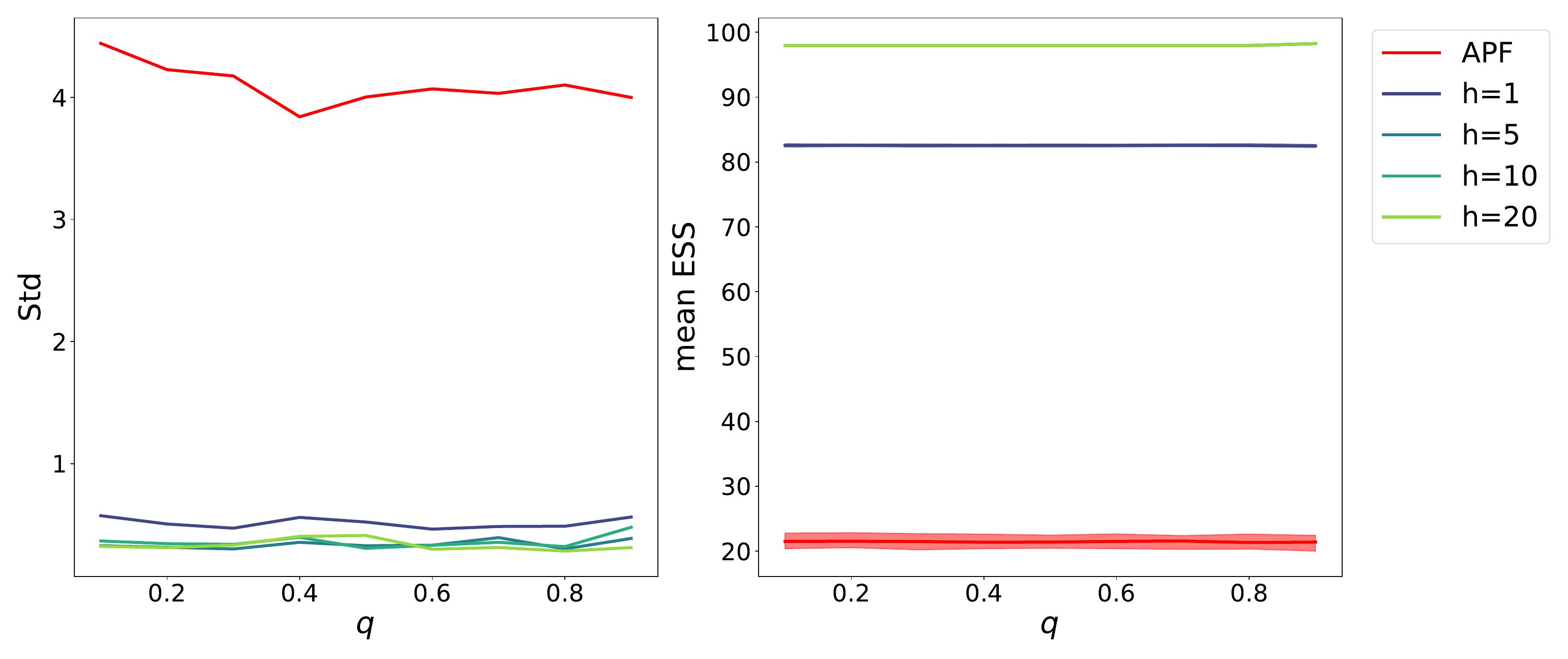}
    \caption{Effective sample size and marginal likelihood standard deviation sensitivity to $\mathbf{q}$. The first column shows the marginal likelihood standard deviation, while the second column reports ESS where mode, $5\%$-quantile, and $95\%$-quantile are reported.}
    \label{fig:q_std_ESS_SIS}
\end{figure}

Outcomes from the SIS model are shown in figure \ref{fig:SIS_data}. The full table on the marginal likelihood standard deviation is reported in table \ref{tab:table_SIS_full}. Complete marginal likelihood contour plot are reported in figure \ref{fig:beta_lambda_SIS} and figure \ref{fig:beta_gamma_SIS} in log scale and normalized to have their max in zero.

We measure the sensitivity to $\mathbf{q}$ of APF and our method when $h=1,5,10,20$ when $P=512$. We study the performance of the considered methods when $\mathbf{q}$ varies, precisely we choose a scenario where data are generated according to $\mathbf{q} = [0.8,0.8]^{\mathrm{T}}$, but the algorithms use $\mathbf{q} = [i, i]^{\mathrm{T}}$ with $i \in [0.1,\dots,0.9]$. As for the previous experiment we compute the standard deviation and ESS bands by running $100$ times algorithm \ref{alg:SMC_general} for each framework. Results are reported in figure \ref{fig:q_std_ESS_SIS}. APF is always associated with a higher standard deviation of the marginal likelihood estimate and to a lower ESS. As for the previous experiments, we have a significant improvement in both standard deviation and ESS when choosing $h=1$, this gets even better when $h \geq 5$, especially from an ESS perspective.

We now infer $(\beta_0,\beta_\lambda, \beta_\gamma,\mathbf{q})$ from data generated from DGP. For this experiment, we employ a Particle marginal Metropolis-Hastings (PMMH) \citep{andrieu2010particle} using an SMC with our proposal distribution and resampling scheme along with the following set of priors: $\mathcal{N}(\beta_0^{(1)}|0, 3)$, $\mathcal{N}(\beta_0^{(2)}|0, 3)$, $\mathcal{N}(\beta_\lambda^{(1)}|0, 3)$, $\mathcal{N}(\beta_\lambda^{(2)}|0, 3)$, $\mathcal{N}(\beta_\gamma^{(1)}|0, 3)$, $\mathcal{N}(\beta_\gamma^{(1)}|0, 3)$, $\mathcal{U}nif(\mathbf{q}^{(1)}|0,1)$, $\mathcal{U}nif(\mathbf{q}^{(2)}|0,1)$. As proposal distribution for the parameters, we use a Gaussian random walk on the $\log$-parameters with variance chosen to match the optimal acceptance rate of $23 \%$.

\begin{figure}[httb!]
    \centering
    \includegraphics[width=\textwidth]{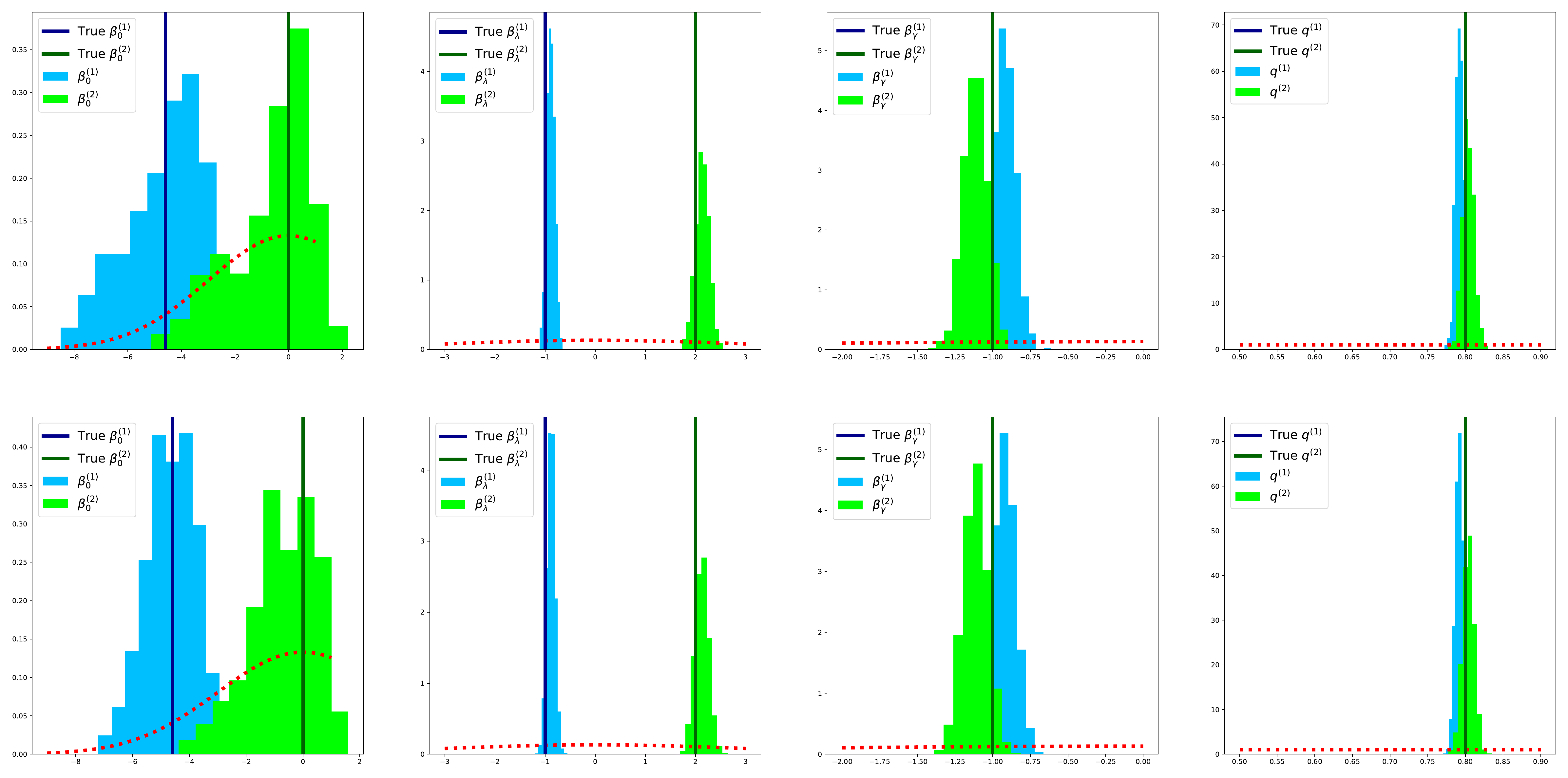}
    \caption{Posterior distribution over the parameters of interest. On the columns $(\beta_0,\beta_\lambda, \beta_\gamma,\mathbf{q})$ from left to right. On the rows $h=5,10$ from top to bottom. Priors are reported in red dotted lines. Vertical lines are the DGP. }
    \label{fig:PMMH_SIS_posteriors}
\end{figure}

\begin{figure}[httb!]
    \centering
    \includegraphics[width=\textwidth]{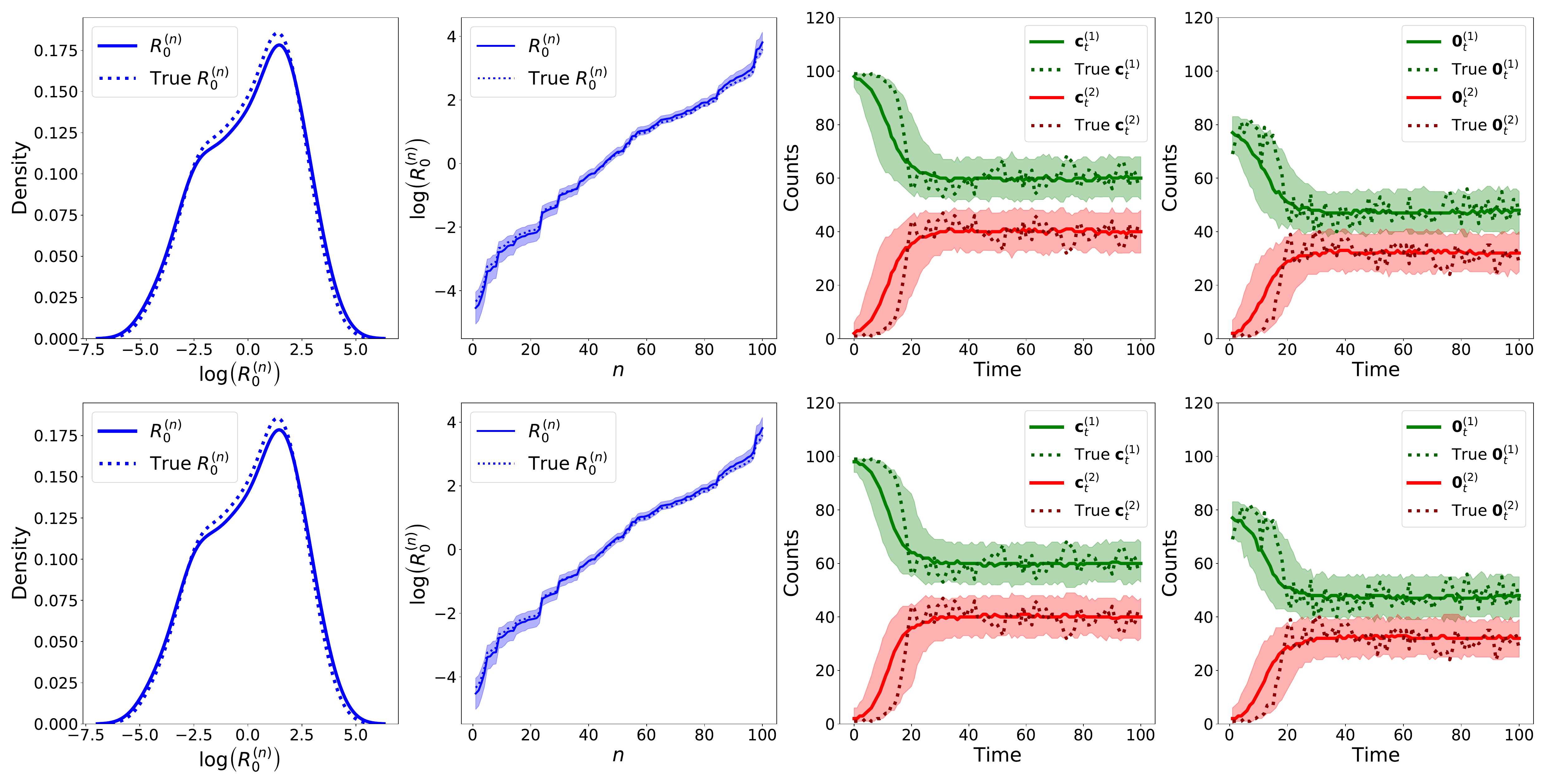}
    \caption{$R_0$ distribution and posterior predictive. First column distribution of the mean $R_0^{(n)}$. Second column  $R_0^{(n)}$ and credible bands for each individual (sorted for increasing $R_0^{(n)}$). The third and fourth columns are posterior predictive over $\mathbf{x}_t$ and $\mathbf{y}_t$. Rows from top to bottom $h=5,10$.}
    \label{fig:PMMH_SIS_R0}
\end{figure}

We run the PMMH for $100000$ iterations and we used a burn-in period of $10000$ and thinning. Marginal posterior distributions are reported in figure \ref{fig:PMMH_SIS_posteriors}, we can notice that posterior are ``peaky'' and close to the DGP, with the exception of $\beta_0$ which has a flatter posterior, which is more difficult to learn given that the only information is derived from $\mathbf{y}_1$. To push our study one step further we also plot $R_0^{(n)}$ distribution and posterior predictive in figure \ref{fig:PMMH_SIS_R0}. We can observe that the PMMH learned the $R_0^{(n)}$ distribution on a global level (first column) and on an individual level (second column), with $R_0^{(n)}$ from the DGP almost indistinguishable. For the posterior predictive we sample $200$ times from the posterior and we then simulate $200$ epidemics with those parameters. Figure \ref{fig:PMMH_SIS_R0} shows good coverage of the aggregated data (fourth column) and of the aggregated latent data (third column). The latter is not available during the inference process, but it is stored at simulation time to add an additional level of comparison.

\subsection{Susceptible-exposed-infected-removed}

\begin{figure}[httb!]
    \centering
    \includegraphics[width =\textwidth]{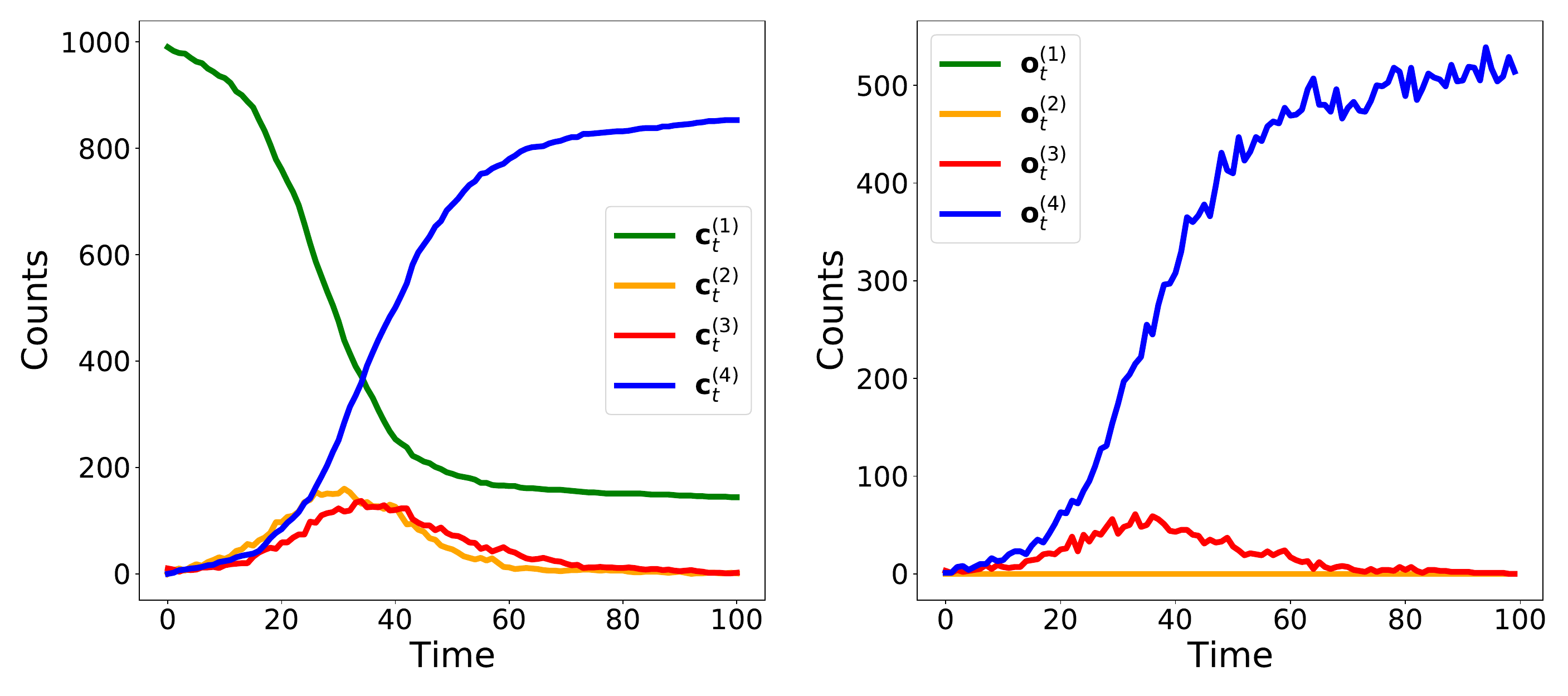}
    \caption{SEIR epidemic, on the left the unobserved process, on the right the observed data.}
    \label{fig:SEIR_data}
\end{figure}

\begin{table}[httb!]
    \centering
    \resizebox{11cm}{!}{
    \begin{tabular}{l|ll|ll|ll|ll}
    \hhline{~|--|--|--|--}
          & h=5   & 0.9s   & h=10   & 3.5s   & h=20   & 5.45s   & h=50   & 9.03s   \\
    \hhline{~|--|--|--|--}
          & DGP   & NDGP   & DGP    & NDGP   & DGP    & NDGP    & DGP    & NDGP    \\
    \hline
     P    & std   & std    & std    & std    & std    & std     & std    & std     \\
     64   & 49.1  & 29.59  & 23.36  & 34.47  & 8.44   & 18.13   & 8.48   & 13.93   \\
     128  & 58.18 & 68.2   & 20.47  & 32.6   & 9.59   & 18.32   & 6.93   & 11.71   \\
     256  & 43.99 & 72.48  & 21.3   & 31.35  & 7.56   & 16.07   & 6.77   & 11.16   \\
     512  & 48.23 & 74.78  & 18.37  & 28.64  & 6.39   & 15.76   & 6.23   & 10.72   \\
     1024 & 45.64 & 66.12  & 19.36  & 29.75  & 5.71   & 14.1    & 5.38   & 10.5    \\
     2048 & 42.7  & 58.37  & 15.03  & 24.68  & 5.69   & 13.25   & 4.57   & 10.45   \\
    \hline
    \end{tabular}
    }
    \caption{Table reporting standard deviation for our method when $h=5,10,20$ under DGP and NDGP with $P=64,128,256,512,1024,2048$. The mean computational cost is reported in the first row with the name of the algorithm.}
    \label{tab:table_SEIR_full}
\end{table}

We can make the SEIR model heterogeneous by considering $d \in \mathbb{N}$ and by defining $(\mathbf{w}_n)_{n \in [1:N]}$ as the collection of $d$-dimensional vectors collecting the individual-specific covariates. From $\mathbf{w}_n$ we can define the initial distribution:
\begin{equation*}
	\mathbf{p}_{n,0} = 
	\begin{bmatrix}
	1-\frac{1}{1+\exp{(-\beta_0^{\mathrm{T}} w_n)}} \\
	0\\
	\frac{1}{1+\exp{(-\beta_0^{\mathrm{T}} w_n)}} \\
	0
	\end{bmatrix} \text{ for } n \in [1:N] \text{ and } \beta_0 \in \mathbb{R}^{d}
\end{equation*}
and calculate $(\mathbf{K}_{n,\bullet})_{n \in [1:N]}$ as:
$$
\mathbf{K}_{n,c} = 
\begin{bmatrix}
\small
1-\frac{1}{1+\exp{(-\beta_{\lambda}^{\mathrm{T}} w_n)}}\frac{\mathbf{c}^{(3)}}{N} &   \frac{1}{1+\exp{(-\beta_{\lambda}^{\mathrm{T}} w_n)}}\frac{\mathbf{c}^{(3)}}{N} & 0 & 0\\
0 & \exp(-\rho) & 1-\exp(-\rho) & 0\\
  0 & 0 & 1-\frac{1}{1+\exp{(-\beta_{\gamma }^{\mathrm{T}} w_n)}} & \frac{1}{1+\exp{(-\beta_{\gamma }^{\mathrm{T}} w_n)}}\\
  0 & 0 & 0 & 1
\end{bmatrix}
$$    
for $n \in [1:N]$ and with $\beta_\lambda,\beta_\gamma \in \mathbb{R}^{d}$ and $\rho \in \mathbb{R}_+$. In this model, we have individual-specific probabilities of infection and recovery and a homogeneous latent period of $1 \slash \rho$. Outcomes from the model are shown in figure \ref{fig:SEIR_data}.

\begin{figure}[httb!]
    \centering
    \includegraphics[width = \textwidth]{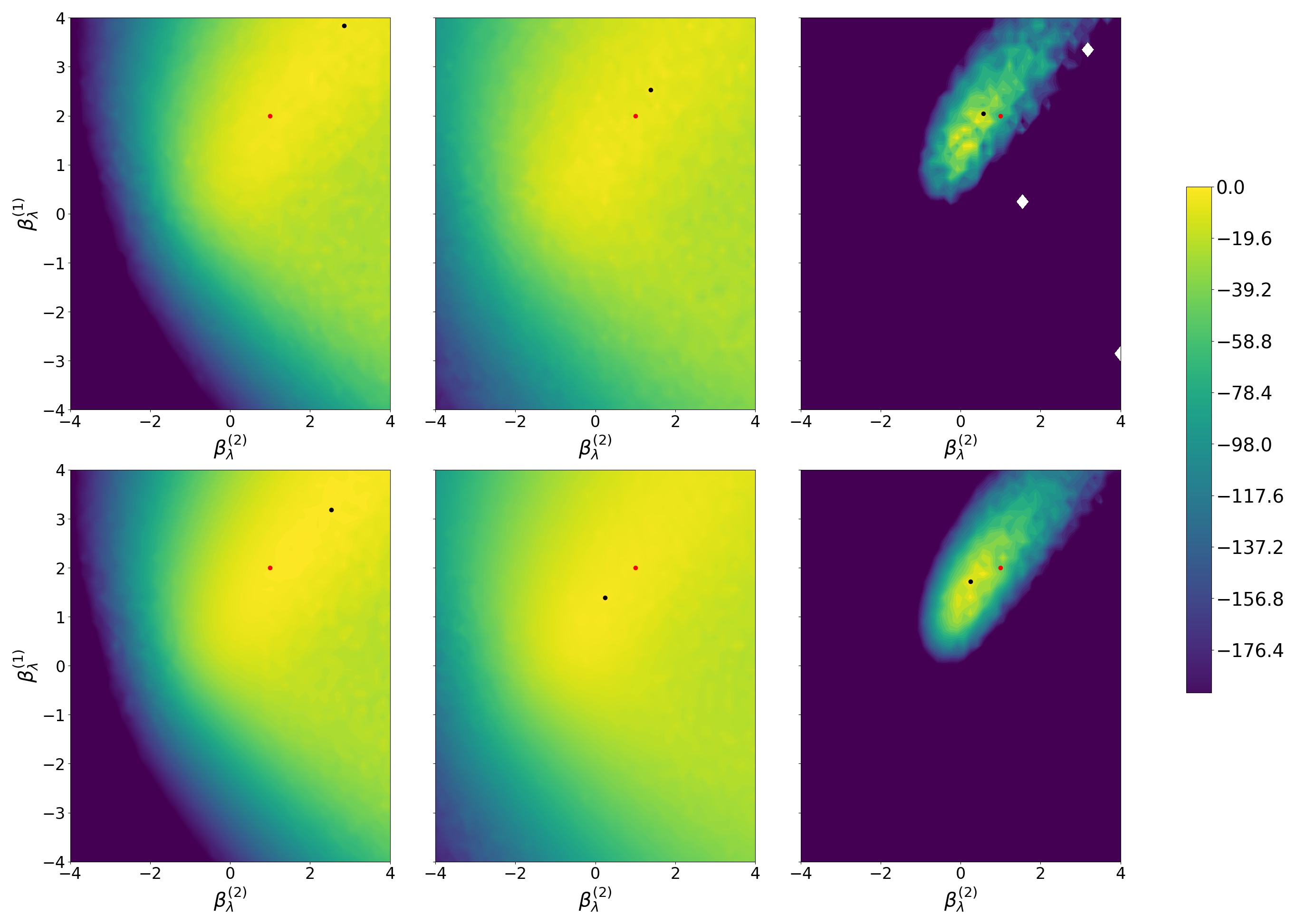}
    \caption{Marginal likelihood contour plot on a $\beta_\lambda$ grid in log-scale. Columns refer to $t=25, 50, 100$ from left to right. Rows refer to $h=10,20$ from top to bottom. The colorbar is common across the plots and in each plot, the maximum is set to $0$. In red is the DGP and in black is the MLE on the grid.}
    \label{fig:beta_lambda_SEIR}
\end{figure}

\begin{figure}[httb!]
    \centering
    \includegraphics[width = \textwidth]{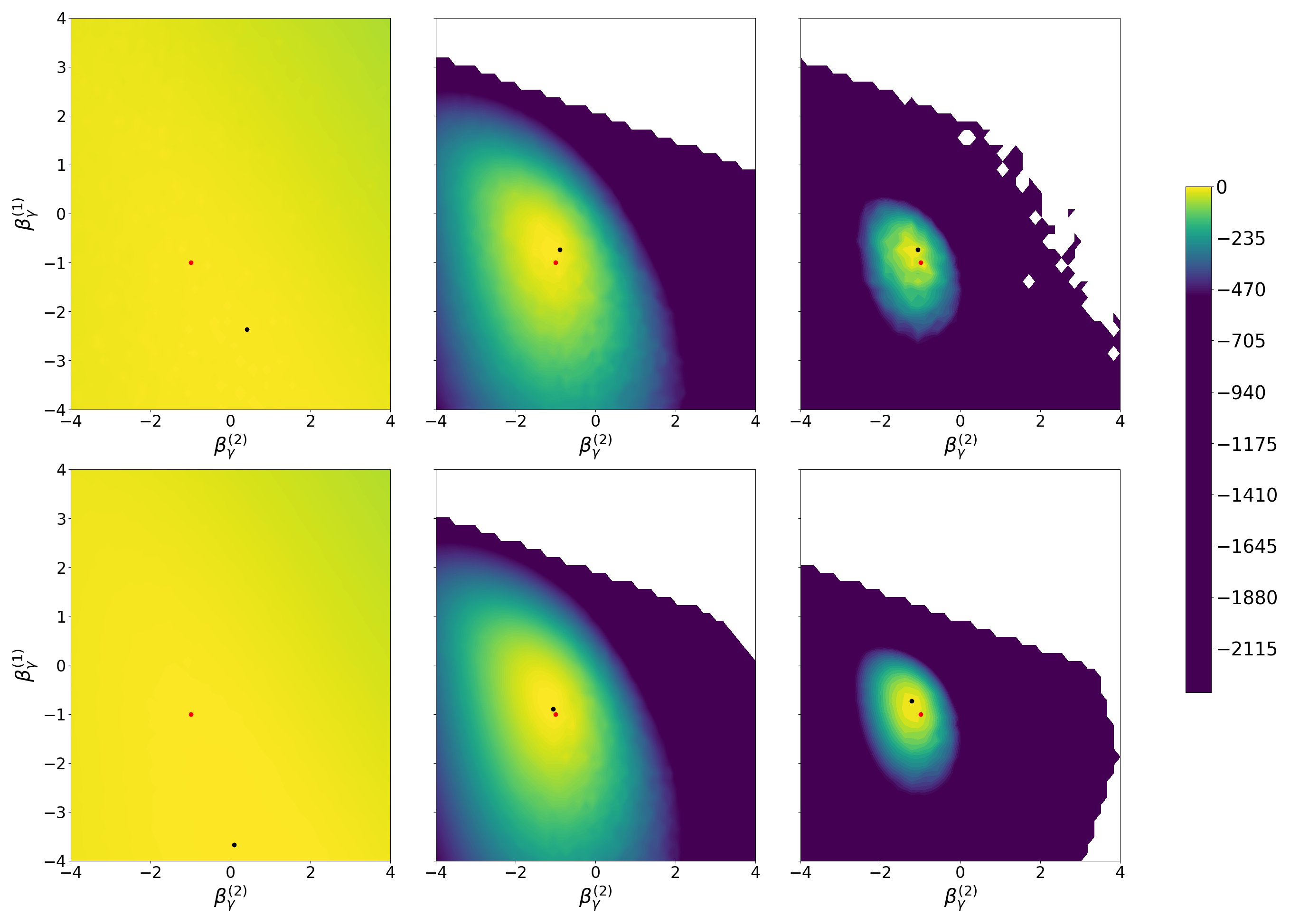}
    \caption{Marginal likelihood contour plot on a $\beta_\gamma$ grid in log-scale. Columns refer to $t=25, 50, 100$ from left to right. Rows refer to $h=10,20$ from top to bottom. The colorbar is common across the plots and in each plot, the maximum is set to $0$. In red is the DGP and in black is the MLE on the grid.}
    \label{fig:beta_gama_SEIR}
\end{figure}

\begin{figure}[httb!]
    \centering
    \includegraphics[width = 0.8\textwidth]{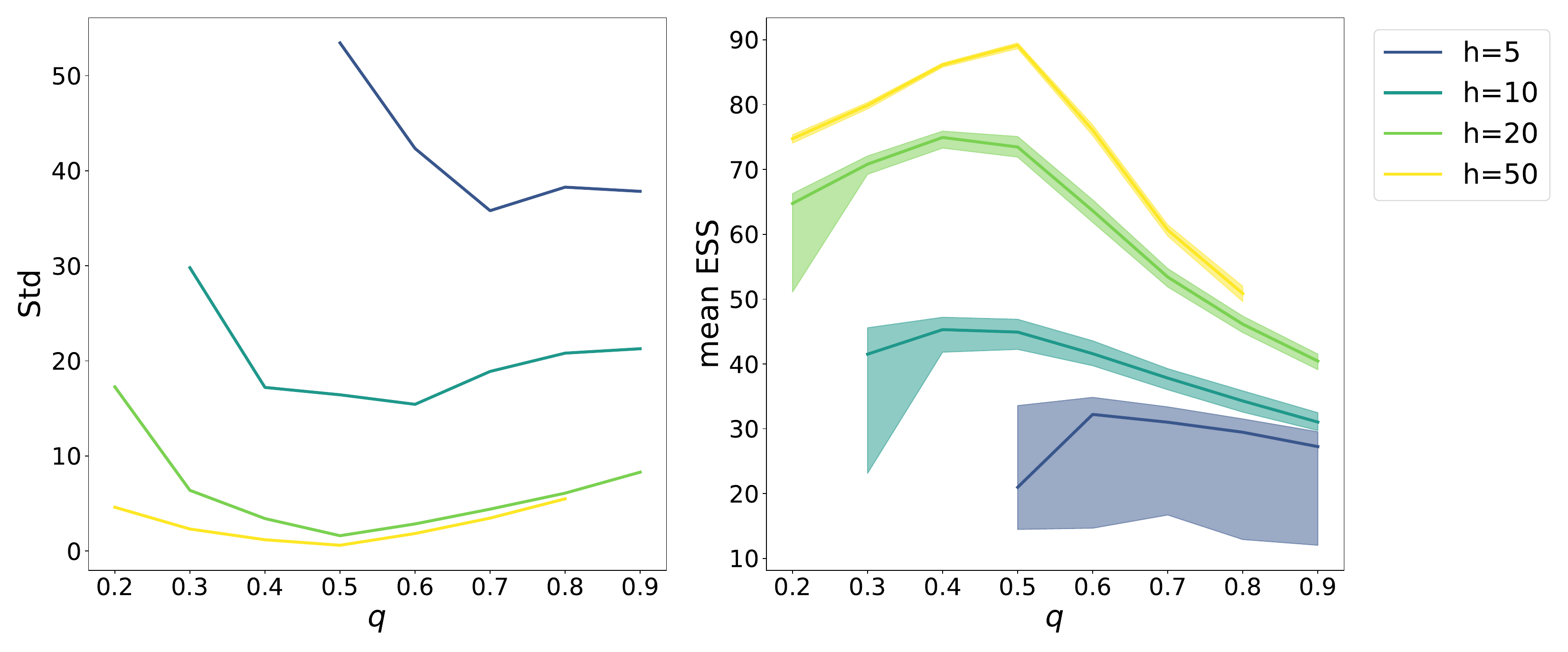}
    \caption{Effective sample size and marginal likelihood standard deviation sensitivity to $\mathbf{q}$. The first column shows the marginal likelihood standard deviation, while the second column reports ESS where mode, $5\%$-quantile, and $95\%$-quantile are reported.}
    \label{fig:q_std_ESS_SEIR_NDGP}
\end{figure}

The full table on the marginal likelihood standard deviation is reported in table \ref{tab:table_SEIR_full}. Complete marginal likelihood contour plots are reported in figure \ref{fig:beta_lambda_SEIR} and figure \ref{fig:beta_gama_SEIR} in log scale and normalized to have their max in zero.

Sensitivity to $\mathbf{q}$ is measured as for the SIS case, with data generated with $\mathbf{q} = [0,0,0.4,0.6]^{\mathrm{T}}$ and our algorithm is run using $P=512$ and $\mathbf{q} = [0, 0, i, i]^{\mathrm{T}}$ with $i \in [0.1,\dots,0.9]$. Figure \ref{fig:q_std_ESS_SEIR_NDGP} reports the results for different choices of $h$. As expected an increase in $h$ is associated with a smaller standard deviation and a bigger mean ESS. It is also important to mention small values of $h$ are more likely to fail even if they are close to the DGP, see $h=5$, while higher values of $h$ report $-\infty$ when they are far from the DGP, see $h=50$ on the values $0.1,0.9$.


\vskip .65cm
\noindent
Department of Mathematics and Statistics, Lancaster University, UK
\vskip 2pt
\noindent
E-mail: l.rimella@lancaster.ac.uk
\vskip 2pt

\noindent
Department of Mathematics and Statistics, Lancaster University, UK
\vskip 2pt
\noindent
E-mail: c.jewell@lancaster.ac.uk
\vskip 2pt

\noindent
Department of Mathematics and Statistics, Lancaster University, UK
\vskip 2pt
\noindent
E-mail: p.fearnhead@lancaster.ac.uk
\vskip 2pt

\end{document}